\documentstyle[preprint,eqsecnum,aps]{revtex}
\tighten
\begin{document}
\preprint{UCONN-96-03}
\title{Anomalous Chiral Action from the Path-Integral}
\author{M.M. Islam and S.J. Puglia}
\address{Department of Physics, University of Connecticut, Storrs, CT 06269}
\maketitle
\normalsize

\begin{abstract}

By generalizing the Fujikawa approach, we show in the path-integral
formalism: (1) how the infinitesimal variation of the fermion measure can be integrated
to obtain the full  anomalous chiral action; (2) how the action derived in this
way can be identified as the Chern-Simons term in five dimensions, if the anomaly is consistent; 
(3) how the regularization can be carried out, so as to lead to the consistent
anomaly and not to the covariant anomaly. Our method uses  Schwinger's
``proper-time'' representation of the Green's function and the gauge invariant
point-splitting technique. We find that the consistency requirement and the point-splitting
technique allow both an anomalous and a non-anomalous action. In the end, the nature of the
vacuum determines whether we have an anomalous theory, or, a non-anomalous theory.
\end{abstract}
\section{Introduction}

It has become increasingly evident that the anomalous action in chiral
theories incorporates important nonperturbative dynamics. For example, the
Wess-Zumino-Witten (WZW) action in the gauged non-linear $\sigma $-model has
been central to the development of the topological soliton model of the
nucleon [1,2]. The predictions of the WZW action that (1) the baryonic current is
topological, and (2) the vector meson $\omega $ couples to it as a gauge boson
have been crucial in realizing that high-energy
elastic $pp$ and $\bar pp$ scattering, in fact, provide strong evidence in
favor of the soliton model of the nucleon [3]. These developments have
motivated us to inquire into the quantum field theory origin of the
nonperturbative anomalous action in the context of the path-integral
formalism. We note that the original geometric argument of Witten in
writing down the anomalous action was based on an intuitive analogy with the
Dirac monopole quantization and the recognition that $\pi _5\left(
SU(3)\right)=Z $ provided a winding number for the mapping of a five
dimensional space onto the internal space of $SU(3)$ [4,5]. In the
differential geometric approach that followed [6,7,8], one identified the
anomalous action as a Chern-Simons (CS) term in a five dimensional space and
showed that it automatically satisfied the Wess-Zumino (WZ) consistency
condition required of the non-Abelian anomaly.

We have approached the question of obtaining the anomalous action in a chiral theory in
a different way. The work of Fujikawa [9,10] suggests that, in the
path-integral formalism, the chiral anomaly originates from the variation
of the fermion measure under an infinitesimal chiral gauge transformation.
Therefore, in principle, one may integrate all such variations to determine the full
anomalous action. Investigation of this approach [11], however, points to
two major difficulties: (1) there is no simple way of seeing how the full
anomalous action becomes a five dimensional quantiy from the integration of
the four dimensional variations; (2) how Fujikawa's regularization method [9]
can be modified so that, instead of the covariant anomaly, one obtains the
consistent anomaly [12]. Fujikawa's approach, however, has the appealing
feature that the anomalies are contained in the path-integral formalism
from the very beginning via Jacobians, and do not have to be brought in
through additional geometric considerations.

Against the above background, we have addressed the following questions:

1. Is it possible to obtain the variation of the fermion measure in such a
way that the origin of the chiral anomaly can be easily seen, and at the
same time the variations can be integrated to lead to the full five-dimensional 
anomalous action?

2. Can it be established that the action obtained above is, in fact, a
Chern-Simons term as given by the differential geometric approach ?

3. Is it possible to determine how the regularization should be carried out,
so that the anomaly is consistent (as required by the WZ consistency condition)? 

In this paper, we present the results of our investigation of the above
questions. We would like to mention that even though very technical and
extensive studies on the subject of the chiral action exist [13],
the kind of generalization of the Fujikawa approach we envisage has not
been done before.
\section{Anomalous Action from the Fermion Measure}

We consider a chiral theory of a massless left-handed fermion
that interacts via a non-Abelian gauge field $A_\mu(x)$. The
path-integral representation of the action functional (or, effective action) $W[A]$ of
the theory in Euclidean space is
\begin{equation}
e^{iW[A]}=%
\frac{1}{{\cal N}}\int d\psi_Ld\bar{\psi}_Le^{\int d^4x\bar{\psi}_L(x)\gamma^\mu %
(\partial_\mu +A_\mu(x))\psi_L(x)}\ \ . \label{2.1}
\end{equation}
Here, $A_\mu = -iA^a_\mu T^a$ and the $T^a$'s are the generators
of the symmetry group. $\gamma^\mu$'s are antihermitian,
$(\gamma^\mu)^{\dag} = -\gamma^\mu$ $(\mu = 1,2,3,4)$; $\bar{\psi}_L=\psi_L^{\dag}\gamma^4=\psi_L^{\dag} i\gamma^0
$. 
The Euclidean metric is $g_{11}=g_{22}=g_{33}=g_{44}=-1$. We first
consider extending the gauge field $A_\mu(x)$ from the
four-dimensional space to a five-dimensional space in the
following way. The fifth dimension is a parameter space with $t$
denoting the parameter and
\begin{equation}
{\cal A}_\mu(x,t) = U^{-1}(x,t)A_\mu(x)U(x,t)+U^{-1}(x,t)\partial_\mu %
U(x,t), \label{2.2}
\end{equation}
where $U(x,t)$ is a unitary matrix: $U(x,t)=e^{\theta (x,t)}$, 
$\theta(x,t)=-iT^a\theta^a(x,t)$ ($T^a$'s are hermitian
matrices; $tr\ T^aT^b=\frac{1}{2}\delta^{ab}$, $\left[ %
T^a,T^b\right]=if^{abc}T^c$). To obtain the effective action
$W[A]$, we adopt the following strategy.

We define first a parameter dependent effective action $W[{\cal%
A}(x,t)]$
\begin{equation}
e^{iW[{\cal A}(x,t)]}=
\frac{1}{{\cal N}}\int d\psi_Ld\bar{\psi}_Le^{\int d^4x\bar{\psi}_L(x)\gamma^\mu %
(\partial_\mu +{\cal A}_\mu(x,t))\psi_L(x)} \label{2.3}
\end{equation}
and require $U(x,t)$ to satisfy the boundary conditions:
\begin{equation}
U(x,t)=\left\{\begin{array}{ll}
1, & t=0 \\
U(x), & t=1.
\end{array}
\right.  \label{2.4}
\end{equation}
$U(x)$ represents a unitary field that transforms under a
gauge transformation in the following way:
\begin{equation}
U(x)\stackrel{g}{\rightarrow}\ U^g(x)=g^{-1}(x)U(x). %
\label{2.5}
\end{equation}
Since under a gauge transformation
\begin{equation}
A_\mu(x)\stackrel{g}{\rightarrow}\ A^g_\mu= %
g^{-1}(x)A_\mu (x)g(x)+g^{-1}(x)\partial_\mu g(x)\ \ , \label{2.6}
\end{equation}
we find that 
\begin{equation}
A^U_\mu(x)=U^{-1}(x)A_\mu (x)U(x)+U^{-1}(x)\partial_\mu U(x) %
\label{2.7}
\end{equation}
is gauge invariant. From the boundary conditions (\ref{2.4}), we
see that at $t=0$ $W[{\cal A}(x,t)]$ coincides with $W[A]$, and
at $t=1$ it becomes a gauge invariant effective action $W[A^U]$. From
(\ref{2.3}), we have
\begin{eqnarray}
e^{iW[{\cal A}(x,t+\delta t)]}&=&\frac{1}{{\cal N}}\int d\psi_Ld\bar{\psi}_Le^{\int d^4\!x\bar{\psi}_L(x)\gamma^\mu%
(\partial_\mu +{\cal A}_\mu(x,t+\delta t))\psi_L(x)} \nonumber \\
&=&\frac{1}{{\cal N}}\int d\psi_Ld\bar{\psi}_Le^{\int d^4x\bar{\psi}_L(x)
U^{-1}(x,t+\delta t)\gamma^\mu(\partial_\mu + A_\mu(x))U(x,t+\delta t)\psi_L(x)} \nonumber \\
&=&\frac{1}{{\cal N}}\int d\psi_Ld\bar{\psi}_Le^{\int d^4\!x\bar{\psi}_L(x)
[1-U^{-1}\partial_tU\delta t]\gamma^\mu(\partial_\mu +{\cal  A}_\mu(x,t))[1+U^{-1}\partial_tU\delta t]\psi_L(x)} 
\label{2.8}
\end{eqnarray}
We now change the Grassmann integration variables $\psi_L,\ %
\bar{\psi}_L$ to $\psi^{\prime}_L,\ \bar{\psi}^{\prime}_L$:
\begin{equation}
\psi^{\prime}_L=(1+U^{-1}\partial_t U\delta t)\psi_L\ \ ,%
\label{2.9}
\end{equation}
\begin{equation}
\bar{\psi}^{\prime}_L=\bar{\psi}_L(1-U^{-1}\partial_t U\delta%
t)\ . \label{2.10}
\end{equation}
Correspondingly, the fermion measure $d\psi_L d\bar{\psi}_L$
transforms into a new measure $d\psi^{\prime}_L %
d\bar{\psi}^{\prime}_L$ The Jacobian of the transformation can
be calculated following Fujikawa's approach [9,10].

We introduce a complete set of eigenfunctions of the Dirac
operator $\gamma^\mu\left(\partial_\mu+{\cal A}_\mu (x,t)%
\right)$:
\begin{equation}
\gamma^\mu\left(\partial_\mu + {\cal A}_\mu (x,t)\right)
\varphi_n(x,t)=\lambda_n(t)\varphi_n(x,t). \label{2.11}
\end{equation}
From the $\varphi_n$'s, we construct a complete set of
chiral wavefunctions $\varphi_{nC}(x,t)$'s ($C=L,R$):
\begin{eqnarray}
\gamma^5\varphi_{nL}(x,t)&=&-\varphi_{nL}(x,t)\ , \nonumber \\
\gamma^5\varphi_{nR}(x,t)&=&\ \varphi_{nR}(x,t)\ .
\label{2.12}
\end{eqnarray}
$\psi_L(x)$ can now be expanded in terms of the left-handed
wavefunctions:
\begin{equation}
\psi_L=\sum_n a_{nL}\ \varphi_{nL}(x,t)\ , \label{2.13}
\end{equation}
where the $a_{nL}$'s are Grassmann numbers. Similarly, 
$\bar{\psi}_L(x)$ can be expanded in terms of the right-handed
wave functions:
\begin{equation}
\bar{\psi}_L(x)=\sum_n \bar{a}_{nR}\varphi^{\dag}_{nR}(x,t)\ , %
\label{2.14}
\end{equation}
where $\bar{a}_{nR}$'s are another set of Grassmann numbers. The
expansions (\ref{2.13}) and (\ref{2.14}) define
the fermion measure
\begin{equation}
d\psi_Ld\bar{\psi}_L=\prod_n da_{nL}\prod_n d\bar{a}_{nR}%
\label{2.15}
\end{equation}
The Grassmann elements $\psi^{\prime}_L$
and $\bar{\psi}^{\prime}_L$ can also be expanded as in
(\ref{2.13}) and (\ref{2.14}), and the expansions define the
measure
\begin{equation}
d\psi^{\prime}_Ld\bar{\psi}^{\prime}_L=
\prod_n da^{\prime}_{nL} \prod_n d\bar{a}^{\prime}_{nR}. %
\label{2.16}
\end{equation}
Eqs. (\ref{2.9}) and (\ref{2.10}) give us
\begin{equation}
a_{nL}^{\prime}=\sum_m \left(\varphi_{nL},[1+U^{-1}\partial_t
U\delta t]\varphi_{mL}\right)a_{mL} \ \ ,  \label{2.17}
\end{equation}
\begin{equation}
\bar{a}^{\prime}_{nR}=\sum_m \bar{a}_{mR}
\left(\varphi_{mR},[1-U^{-1}\partial_t U\delta t]
\varphi_{nR}\right)\ \ . \label{2.18}
\end{equation}
These relations yield
\begin{eqnarray}
\prod_n da_{nL}\prod_n d\bar{a}_{nR}&=& 
Det\langle\varphi_C|\left[(1+U^{-1}\partial_tU\delta t)a_{-}
+a_{+}\right]
|\varphi_{C}\rangle \nonumber \\ 
& & \times Det \langle\varphi_C|\left[(1-U^{-1}\partial_tU\delta
t)a_{+}+a_{-}\right]|\varphi_{C}\rangle
\prod_n da^{\prime}_{nL}\prod_n d\bar{a}^{\prime}_{nR},
\label{2.19}
\end{eqnarray}
where the determinants are defined with the $\varphi_{nC}$'s as
the basis functions. $a_{-},\ a_{+}$ are the chiral projection
operators: $a_{-}=\frac{1}{2}(1-\gamma^5),\ a_{+}=\frac{1}{2}(%
1+\gamma^5)$, and
\[
\begin{array}{lllllllll}
a_{-}\varphi_{nC} &=&\varphi_{nL} &(C=L)&,&a_{+}\varphi_{nC}&=&0%
&(C=L) \\
&=& 0 &(C=R)&,& &=&\varphi_{nR}&(C=R)
\end{array}
\]
Since $\delta t$ is infinitesimal and $a_{-}+a_{+}=1$, the
determiants in (\ref{2.19}) differ from unity only by
infinitesimal variations. Eq. (\ref{2.19}) can then be written as
\begin{eqnarray}
d\psi_L d\bar{\psi}_L &=&e^{Tr\langle\varphi_C|U^{-1}\partial_tUa_{-}|\varphi_{C}\rangle\delta t%
-Tr\langle\varphi_C|U^{-1}\partial_tUa_{+}|\varphi_{C}\rangle\delta t} 
\  d\psi^{\prime}_L d\bar{\psi}_L^{\prime} \nonumber \\
&=&e^{-Tr\langle\varphi_C|U^{-1}\partial_tU\gamma^5|\varphi_{C}\rangle\delta t}
\ d\psi^{\prime}_L d\bar{\psi}_L^{\prime} \ \ ,
\label{2.20}
\end{eqnarray}
where $Tr$ stands for an integral over the coordinate space and
trace over the group and Dirac matrices $(Tr=\int d^4\! x\  tr=\int
 d^4\! x\ tr_G tr_D)$. Switching back to the eigenfunctions 
 $\varphi_n(x,t)$'s as basis functions, Eq. (\ref{2.20})
 becomes
 \begin{equation}
 d\psi_L d\bar{\psi}_L= e^{-\int d^4\!x\  tr\left[U^{-1}\partial_tU\gamma^5 %
 \sum_n\varphi_n(x,t)\varphi_n^{\dag}(x,t)\right]\delta t}
 d\psi^{\prime}_L d\bar{\psi}_L^{\prime}\ \ .%
 \label{2.21}
 \end{equation}
 From (\ref{2.8}) and (\ref{2.21}), we obtain
 \begin{eqnarray}
 e^{iW[{\cal A}(x,t+\delta t)]}&=&e^{-\int d^4\!x\  tr\left[U^{-1}\partial_tU\gamma^5%
 \sum_n\varphi_n(x,t)\varphi_n^{\dag}(x,t)\right]\delta t}  \nonumber \\
& &\times\frac{1}{{\cal N}} \int d\psi^{\prime}_L d\bar{\psi}_L^{\prime}e^{\int d^4\!x\ \bar{\psi}^{\prime}_L(x)\gamma^\mu%
(\partial_\mu +{\cal A}_\mu(x,t))\psi^{\prime}_L(x)} \nonumber  \\
&=& e^{-\int d^4\!x\  tr\left[U^{-1}\partial_tU\gamma^5\sum_n\varphi_n(x,t)\varphi_n^{\dag}(x,t)\right]\delta t}%
 e^{iW[{\cal A}(x,t)]}\ \ .
 \label{2.22}
 \end{eqnarray}
 From (\ref{2.22}), we find
 \begin{equation}
 \frac{\partial W[{\cal A}(x,t)]}{\partial t}=
 i\int d^4\! x\ tr\left[U^{-1}(x,t)\partial_t U(x,t)\gamma^5
 \sum_n\varphi_n(x,t)\varphi_n^{\dag}(x,t)\right]\ \ . \label{2.23}
 \end{equation}
 Therefore, 
 \begin{eqnarray}
 W[A]&=&-i\int^1_0 dt\int d^4\! x\ tr\left[U^{-1}(x,t)\partial_t
 U(x,t)\gamma^5\sum_n\varphi_n(x,t)\varphi_n^{\dag}(x,t)\right]
 \nonumber \\
 & & +W[A^U]
 \label{2.24}
 \end{eqnarray}
 As noted earlier, $W[A^U]$ is a gauge invariant effective
 action.

 We can derive an alternative expression for the derivative
 of $W[{\cal A}(x,t)]$ with respect to the parameter $t$ by directly
 differentiating (\ref{2.3}): 
 \begin{eqnarray}
 i \frac{\partial W[{\cal A}(x,t)]}{\partial t}&=& 
\int\langle\bar{\psi}_L\gamma^\mu(-iT^a)\psi_L\rangle
\dot{{\cal A}}_{\mu}^a (x,t)d^4x \nonumber \\
&=&\int j^{\mu ,a}_L(x,t)\dot{{\cal A}}_{\mu}^a (x,t)d^4\! x\ \ , 
\label{2.25}
\end{eqnarray}
where $j^{\mu ,a}_L$ is the chiral current:
\begin{eqnarray}
j^{\mu ,a}_L(x,t) &=& i \frac{\delta W[{\cal A}(x,t)]}{\delta
{\cal A}_{\mu}^a(x,t)} \label{2.26} \\
&=& \langle\bar{\psi}_L\gamma^\mu(-iT^a)\psi_L\rangle.
\label{2.27}
\end{eqnarray}
Eq.(\ref{2.25}) can be written as 
\begin{equation}
\frac{\partial W[{\cal A}(x,t)]}{\partial t}=2i\int tr_G\left[\dot{{\cal A}}_\mu(x,t)j^{\mu}_L(x,t)
\right]d^4\!x, 
\label{2.28}
\end{equation}
where $j^\mu_L=-iT^aj^{\mu ,a}_L$. Now, 
\begin{eqnarray}
\dot{{\cal A}}_\mu (x,t)&=&\partial_t {\cal A}_\mu(x,t) \nonumber \\
&=& D_\mu({\cal A})\left(U^{-1}(x,t)\partial_tU(x,t)\right),
\label{2.29}
\end{eqnarray}
where $D_\mu({\cal A})=\partial_\mu + [{\cal A}_\mu ,\ ]$. Therefore,
\begin{eqnarray}
\frac{\partial W[{\cal A}(x,t)]}{\partial t}&=&2i\int tr_G\left[D_\mu({\cal%
A})\left(U^{-1}(x,t)\partial_tU(x,t)\right)j^{\mu}_L(x,t)\right]d^4\!x \nonumber \\
&=& -2i\int tr_G\left[U^{-1}(x,t)\partial_tU(x,t)D_\mu({\cal A})j^{\mu}_L(x,t)\right]d^4\!x
\label{2.30}
\end{eqnarray}
Comparing Eqs. (\ref{2.23}) and (\ref{2.30}), we obtain
\begin{equation}
D_\mu({\cal A})j^{\mu}_L(x,t)=-\frac{1}{2}tr_D\left[\gamma^5\sum_n\varphi_n(x,t)
\varphi^{\dag}_n(x,t)\right]\ \ . \label{2.31}
\end{equation}
Since the right-hand side (RHS) of Eq. (\ref{2.31}) is not well defined, it has to be regularized
to be meaningful. Let $a({\cal A})$ be the corresponding regularized
expression:
\begin{equation}
a({\cal A})\equiv -\frac{1}{2} tr_D\gamma^5 \sum_n\varphi_n(x,t)\varphi^{\dag}_n(x,t)|_{reg} \label{2.32}
\end{equation}
Then,
\begin{equation}
D_\mu({\cal A})j^{\mu}_L(x,t)=a({\cal A});
\label{2.33}
\end{equation}
that is, $a({\cal A})$ is the covariant divergence of the chiral current
$j^{\mu}_L$. A nonvanishing $a({\cal A})$ means that the current is not conserved covariantly, and
$a({\cal A})$ is referred to as the chiral anomaly. From Eq.(\ref{2.24}), we see that
\begin{equation}
W[A]=\Gamma[A,U]+W[A^U]\ \ ,
\label{2.34}
\end{equation}
where
\begin{equation}
\Gamma[A,U]=2i\int_0^1dt\int d^4\!x\ tr_G\left[U^{-1}(x,t)\partial_tU(x,t)a({\cal A})\right]
\label{2.35}
\end{equation}
can now be identified as the anomalous action (or, WZW action).

We note the following: 1) Eq.(\ref{2.22}) shows how an infinitesimal
change of the parameter $t$ leads to an infinitesimal variation
of the fermion measure; 2) Eq. (\ref{2.31}) makes it evident that such a
variation of the fermion measure can produce the anomaly; 3) Eq.
(\ref{2.35}) shows how such variations can be integrated to obtain the
full anomalous action. Our method of calculating the variation
of the fermion measure follows that of Fujikawa. However, by
extending the guage field $A_{\mu}(x)$ to a parameter space 
(Eq.(\ref{2.2}))and by considering variation of the
parameter , we have been able to generalize the Fujikawa
approach, so as to obtain an anomalous action in five
dimensions.

In the next section (Sec. 3), we prove that, if the anomaly
$a({\cal A})$ is the consistent anomaly, then the anomalous
action $\Gamma [A,U]$ is the Chern-Simons term as given by the
differential geometric approach [6,7]. In sections 4 and 5, we
then investigate how
$\displaystyle tr_D\ \gamma^5\sum_n\varphi_n(x,t) %
\varphi^{\dag}_n(x,t)$ has to be regularized, so that 
$a({\cal A})$, in fact, becomes the consistent anomaly.
\section{ Identifying the Anomalous Action as the Chern-Simons Term }

In this section, our objective is to determine the anomalous action
$\Gamma[A,U]$, assuming that the regularized
anomaly $a({\cal A})$ is the consistent anomaly. The latter assumption implies [12]
\begin{equation}
a({\cal A})=
\tilde{a}({\cal A})-D_\mu({\cal A})X^\mu({\cal A})\ ,
\label{3.1}
\end{equation}
where
\begin{equation}
\tilde{a}({\cal A})=\frac{1}{64\pi^2}\epsilon^{\lambda\mu\rho\sigma}
G_{\lambda\mu}({\cal A})G_{\rho\sigma}({\cal A});
\label{3.2}
\end{equation}
$G_{\lambda\mu}({\cal A})$ is the field tensor: $G_{\lambda\mu}({\cal A})=\partial_\lambda{\cal A}_\mu %
-\partial_\mu{\cal A}_\lambda+[{\cal A}_\lambda,{\cal A}_\mu]$, and $X^\lambda({\cal A})$ is a polynomial
in ${\cal A}$ and its derivatives:
\begin{equation}
X^\lambda({\cal A})=\frac{1}{48\pi^2}\epsilon^{\lambda\mu\rho\sigma}\left[\partial_\mu{\cal A}_\rho{\cal A}_\sigma
+{\cal A}_\sigma\partial_\mu{\cal A}_\rho+\frac{3}{2}{\cal A}_\mu{\cal A}_\rho{\cal A}_\sigma\right]
\label{3.3}
\end{equation}
The first term on the RHS of (\ref{3.1}), $\tilde{a}({\cal A})$, is known as the covariant
anomaly. It is the usual anomaly one obtains using Fujikawa's regularization procedure. Inserting 
(\ref{3.1}) in (\ref{2.35}), and writing ${\cal A}_0(x,t)\equiv U^{-1}(x,t)\partial_t U(x,t)$, we obtain
\begin{eqnarray}
\Gamma[A,U]&=&2i\int_0^1dt\int d^4\!x\ tr\left[{\cal A}_0\left(\tilde{a}({\cal A})-D_\mu({\cal A})X^\mu({\cal A})
\right)\right] \label{3.4} \\
&=&2i\int_0^1dt\int d^4\!x\ tr\left[{\cal A}_0\tilde{a}({\cal A})+\dot{{\cal A}}_\mu X^\mu({\cal A})
\right]\label{3.5} \\
&=& -i\int_0^1dt\int d^4\!x\ \left[{\cal A}^a_0\tilde{a}^a({\cal A})+\dot{{\cal A}}^a_\mu X^{\mu,a}({\cal A})
\right]\ \ ,
\label{3.6}
\end{eqnarray}
where we have used 
$D_\mu({\cal A}){\cal A}_0=\dot{{\cal A}}_\mu$ (Eq.(\ref{2.29}))

The identification of $\Gamma[A,U]$ as the Chern-Simons action now becomes easy,
because of a key observation by Dunne and Trugenberger (DT) [14]. These authors
point out that a Chern-Simons Lagrangian in odd dimensions can be split into two parts,
as in Eq(\ref{3.6}), where the time component of the gauge field is multiplied by
the covariant anomaly, and the time derivative of the space components is multiplied by a
polynomial $X^{\mu ,a}({\cal A})$, whose covariant divergence relates the covariant
anomaly to the consistent anomaly.
In our case, it is of course pertinent to bear in mind that the gauge field ${\cal A}_i(x,t)\ \ (i=\mu,0)$ in the
odd dimensional space has been specified by extending the original gauge field $A_\mu(x)$
via a parameter dependent gauge transformation (Eq.(\ref{2.2})), and the time component ${\cal A}_0$ in
five dimensions has beeen identified as $U^{-1}(x,t)\partial_tU(x,t)$. We now explicitly verify that
$\Gamma[A,U]$ is the Chern-Simons action.

To this end, we start with the definition of the Chern-Simons action in five dimensions [6,7,14]:
\begin{eqnarray}
S_{CS}&=&\frac{i}{8\pi^2}\int_0^1d\xi\ tr\left({\cal A}G_\xi^2\right)\hspace*{3cm}\mbox{(in form notation)} \nonumber \\
&=& \frac{i}{32\pi^2}\int_0^1dt\int d^4\!x\int_0^1d\xi\ tr\left({\cal A}_iG_{jk}(\xi{\cal A})G_{lm}(\xi{\cal A})
\right)\epsilon^{ijklm} \nonumber \\
&=&\frac{i}{32\pi^2}\int_0^1dt\int d^4\!x\ tr\left\{\epsilon^{\mu\nu\rho\sigma}
{\cal A}_0 G_{\mu\nu}({\cal A})G_{\rho\sigma}({\cal A})  \right. \nonumber \\
& & \ \ \ \ \ \ \    +\dot{{\cal A}}_\mu\ 2\int_0^1d\xi\ \xi\left({\cal A}_\nu G_{\rho\sigma}(\xi{\cal A})
+G_{\rho\sigma}(\xi{\cal A}){\cal A}_\nu\right)
\epsilon^{\mu\nu\rho\sigma} \nonumber \\
& & \ \ \ \ \ \ \ \ \ \ \left.-\partial_\mu\left({\cal A}_0\ 2\int_0^1d\xi\ \xi
\left[{\cal A}_\nu G_{\rho\sigma}(\xi{\cal A})+G_{\rho\sigma}(\xi{\cal A}){\cal A}_\nu\right]
\epsilon^{\mu\nu\rho\sigma}\right)\right\}
\label{3.7}
\end{eqnarray}
Let
\begin{equation}
P^{\mu}({\cal A})\equiv2\int_0^1d\xi\ \xi\left[{\cal A}_\nu G_{\rho\sigma}(\xi{\cal A})
+G_{\rho\sigma}(\xi{\cal A}){\cal A}_\nu\right]\epsilon^{\mu\nu\rho\sigma}
\label{3.8}
\end{equation}
Then,
\begin{equation}
P^{\mu}({\cal A})=\frac{4}{3}\left[{\cal A}_\nu\partial_\rho{\cal A}_\sigma
+\partial_\rho{\cal A}_\sigma{\cal A}_\nu+\frac{3}{2}{\cal A}_\nu{\cal A}_\rho{\cal A}_\sigma\right]
\epsilon^{\mu\nu\rho\sigma}
\label{3.9}
\end{equation}
From (\ref{3.7}), we obtain
\begin{equation}
S_{CS}=\frac{i}{32\pi^2}\int_0^1\!\!dt\int\!\!\! d^4\!x\ tr\left\{{\cal A}_0 G_{\mu\nu}({\cal A})G_{\rho\sigma}({\cal A})
\epsilon^{\mu\nu\rho\sigma}+\dot{{\cal A}}_\mu P^{\mu}({\cal A})-\partial_\mu\left({\cal A}_0 P^{\mu}({\cal A})
\right)\right\}
\label{3.10}
\end{equation}
We observe that, if we write $S_{CS}=\int\!dt\ L_{CS}$, then $L_{CS}$ as given by Eq.(\ref{3.10}) is
of the form deduced by DT. Since $\dot{{\cal A}}_\mu =D_\mu({\cal A}){\cal A}_0$ in our case, we get from 
(\ref{3.10})
\begin{eqnarray}
S_{CS}
&=&2i\int_0^1\!\!dt\int\!\! d^4\!x\ tr\left\{U^{-1}\partial_tU\left[\frac{1}{64\pi^2}G_{\mu\nu}({\cal A})
G_{\rho\sigma}({\cal A})\epsilon^{\mu\nu\rho\sigma}
-\frac{1}{64\pi^2}D_\mu({\cal A})P^{\mu}({\cal A})\right]\right\} \nonumber \\
&=&2i\int_0^1\!\!dt\int\!\! d^4\!x\ tr\left\{U^{-1}\partial_tU\left[\tilde{a}({\cal A})-D_\mu({\cal A})X^\mu
({\cal A})\right]\right\} \nonumber \\
&=&\Gamma[A,U]
\label{3.11}
\end{eqnarray}
using Eq.(\ref{3.4}). Establishing this result also implies that---in the path-integral
framework the topological aspects of the theory are contained in the variation of the fermion measure.
\section{Regularization using Proper-Time Representation of Green's Functions }
Our next objective is to specify the regularization of 
${\displaystyle tr\left[\gamma^5\sum_n\varphi_n(x,t)\varphi_{n}^{\dagger}(x,t)\right]}$
To this end, we first introduce the propagator $S(x,y)$ of a massive fermion and a ``proper-time"
integral representation of it:
\begin{eqnarray}
S(x,y) &=& \sum_n \frac{\varphi_n(x)\varphi^{\dagger}_n(y)}{\lambda_n+im}\nonumber \\
&=& (\not\!\!{\cal D}-im)\sum_n \frac{\varphi_n(x)\varphi^{\dagger}_n(y)}{\lambda^2_n+m^2}\nonumber \\
&=& (\not\!\!{\cal D}-im)\int_0^{\infty}\! d\tau e^{-\tau(\not{\cal D}^2+m^2)}\ {\cal M}(x,y) , \label{4.1}
\end{eqnarray}
where $\tau$ is the proper-time variable of Schwinger [15], and
${\cal M}(x,y)\equiv{\displaystyle\sum_n\varphi_n(x)\varphi_{n}^{\dagger}(y)}$ is an infinite dimensional matrix; 
$\not\!\!{\cal D}=\gamma^\mu{\cal D}_\mu,\  {\cal D}_\mu=\partial_\mu+{\cal A}_\mu\ \ (\mu=1-4)$.
For simplicity, we have suppressed the $t$-dependence in (\ref{4.1}). (We do this in other places as well, if
no confusion arises). Eq. (\ref{4.1}) can be expressed in the form
\begin{equation}
S(x,y)=S_R(x){\cal M}(x,y),\label{4.2}
\end{equation}
where
\begin{equation}
S_R(x)=(\not\!\!{\cal D}-im)\int_0^{\infty}\! d\tau \ e^{-\tau(\not{\cal D}^2+m^2)} \label{4.3}
\end{equation}
is the fermion propagator as a differential operator that acts only to the right.
Its inverse is $S^{-1}_R(x)=\not\!\!{\cal D}+im$ (formally, $S_R(x)=\frac{1}{\not{\cal D}+im}$).
Besides (\ref{4.1}), an alternative representation of $S(x,y)$ is
\begin{eqnarray}
S(x,y) &=& \sum_n \frac{\varphi_n(x)\varphi^{\dagger}_n(y)}{\lambda^2_n+m^2}(\stackrel{\leftarrow}{\not %
\!\!{\cal D}}\!-im) \nonumber \\
&=&{\cal M}(x,y)
\int_0^{\infty}\! d\tau \  e^{-\tau(\stackrel{\leftarrow}{\not\!{\cal D}^2}+m^2)}
(\stackrel{\leftarrow}{\not\!\!{\cal D}}\!-im) \ \ ,
\label{4.4}
\end{eqnarray}
where $\stackrel{\leftarrow}{\not\!\!{\cal D}}=\gamma^{\mu}(-\stackrel{\leftarrow}{\partial}_\mu +{\cal A}_\mu(y)),
\mbox{ and }\stackrel{\leftarrow}{\partial_\mu}=\frac{\stackrel{\leftarrow}{\partial}}{\partial y^\mu}$. 
We can also write
\begin{equation}
S(x,y)= {\cal M}(x,y)S_L(y),
\label{4.5}
\end{equation}
where
\begin{equation}
S_L(y)=\int_0^{\infty}\! d\tau \ e^{-\tau(\stackrel{\leftarrow}{\not{\cal D}^2}+m^2)}
(\stackrel{\leftarrow}{\not\!\!{\cal D}}\!-im)
\label{4.6}
\end{equation}
is the fermion propagator as a differential operator that acts to the left. The inverse of $S_L(y)$ is
$S^{-1}_L(y)=(\stackrel{\leftarrow}{\not\!\!{\cal D}}\!+im)$ (formally, 
$S_L(y)=\frac{1}{\stackrel{\leftarrow}{\not{\cal D}}\!+im}$).
Next we  observe that all the following expressions are equivalent to
${\displaystyle tr\left[\gamma^5\sum_n\varphi_n(x)\varphi_{n}^{\dagger}(y)\right]}$:
\begin{eqnarray*}
tr\ \gamma^5\left[S_R{\cal M}(x,y)\right]S^{-1}_L\ ,& tr\ \gamma^5 S_R^{-1}\left[S_R{\cal M}(x,y)\right],& 
tr\ \gamma^5 S_R\left[S^{-1}_R{\cal M}(x,y)\right]\ ,\\ 
tr\ \gamma^5 S^{-1}_R\left[{\cal M}(x,y)S_L\right],& 
tr\ \gamma^5 \left[{\cal M}(x,y)S_L\right]S^{-1}_L\ ,& tr\ \gamma^5\left[{\cal M}(x,y)S^{-1}_L\right]S_L\ \
.
\end{eqnarray*}
For example,
\begin{eqnarray}
tr\ \gamma^5 S_R(x){\cal M}(x,y)S_L^{-1}(y) \nonumber \\
&=& tr\ \gamma^5(\not\!\!{\cal D}-im)\int_0^{\infty}\! d\tau \
e^{-\tau(\not D^2+m^2)}\sum_n\varphi_n(x)\varphi_{n}^{\dagger}(y)
\left(\stackrel{\leftarrow}{\not\!\!{\cal D}}\!+im\right) \nonumber \\
&=& tr\ \gamma^5 \sum_n\varphi_n(x)\varphi_{n}^{\dagger}(y) \ .
\label{4.7}
\end{eqnarray}
Furthermore, we note that each expression involves the fermion propagator and its inverse as differential
operators, and we have above all the six possible arrangements of them acting on the matrix ${\cal M}(x,y)$.

As a first step to obtain a regularized expression for 
${\displaystyle tr\ \gamma^5\sum_n\varphi_n(x,t)\varphi_{n}^{\dagger}(x,t)}$, 
we consider the following average with $x\neq y$  as the
regularized expression for ${\displaystyle tr \ \gamma^5\sum_n\varphi_n(x,t)\varphi_{n}^{\dagger}(y,t)}$:
\begin{eqnarray}
\lefteqn{tr\ \gamma^5\sum_n\varphi_n(x,t)\varphi_{n}^{\dagger}(y,t)|_{reg}} \nonumber \\
 & &=\ \int_{0}^{1}dg\frac{\partial}{\partial g}
 tr\ \gamma^5\sum_n\varphi^{g}_n(x,t)\varphi_{n}^{g\ \dagger}(y,t)\nonumber \\
& &=\  \int^{1}_{0}dg\frac{\partial}{\partial g}\ \frac{1}{6}tr\ \gamma^5\left\{[S_RI(x,y)]S_L^{-1}
+S_R^{-1}[S_RI(x,y)]+
S_R[S_R^{-1}I(x,y)]\right. \nonumber \\
& &\hspace*{3cm}\left. +S_R^{-1}[I(x,y)S_L]+[I(x,y)S_L]S_L^{-1}+[I(x,y)S_L^{-1}]S_L\right\} .
\label{4.8}
\end{eqnarray}
In (\ref{4.8}), $g$ is the gauge coupling constant, $\varphi^g_n$'s are the eigenfunctions of the
Dirac operator $\gamma^\mu(\partial_\mu+g{\cal A}_\mu)$, and we have subtracted out
the free-field term corresponding to
$g=0$ [16] . Furthermore, in (\ref{4.8}) 
${\cal M}(x,y)\equiv{\displaystyle\sum_n\varphi_n(x)\varphi_{n}^{\dagger}(y)}$ has been replaced
by $I(x,y)={\displaystyle\int\frac{d^4k}{(2\pi)^4}e^{ik\cdot(x-y)}}$, i.e., the completeness of the single-particle
wavefunctions has been replaced by that of the plane wave solutions.
A few comments are in order here: 1) If we replace in (\ref{4.8}) $I(x,y)$ by ${\cal M}(x,y)$, we get
back the formal expression: 
$tr\ \gamma^5{\displaystyle \sum_n\varphi_n(x,t)\varphi_{n}^{\dagger}(y,t)}-tr\ \gamma^5 %
{\displaystyle\sum_n\varphi^{0}_n(x,t)\varphi_{n}^{0\ \dagger}(y,t)}$. 2) The form of Eq.(\ref{4.8})
is dictated by the requirement that the matrix structure be symmetric between the variables $x$ and $y$.
3) The particular averaging considered in (\ref{4.8}) corresponds to taking the identity matrix
${\cal M}(x,y)=\displaystyle{\sum_n\varphi_n(x)\varphi_n^{\dag}(y)}$ as the product of the particle propagator
and its inverse (taken in all possible manner).
At this point, the averaging has to be regarded as ad hoc   
and can only be justified a posteriori by showing that it yields the consistent anomaly, and in turn,
the Chern-Simons term with the right normalization.

Introducing the Green's functions $G_R(x,y)={\displaystyle\int_0^1d\tau e^{-\tau(\not{\cal D}^2+m^2)}\ I(x,y)}$ and
$G_L(x,y)={\displaystyle\int_0^1d\tau\  I(x,y)\ e^{-\tau(\stackrel{\leftarrow}{\not{\cal D}^2}+m^2)}}$ [17], we can
express Eq.(\ref{4.8}) in the following form:
\begin{eqnarray}
\lefteqn{tr\ \gamma^5\sum_n\varphi_n(x,t)\varphi_{n}^{\dagger}(y,t)|_{reg}}  \nonumber \\
& &=\ \frac{1}{6}\int_{0}^{1}dg\frac{\partial}{\partial g}\left\{tr\ \gamma^5(\not\!\!{\cal D}-im)G_R(x,y)
(\stackrel{\leftarrow}{\not\!\!{\cal D}}+im)\right.\nonumber \\
& & \hspace*{3cm}+\left.tr\ \gamma^5(\not\!\!{\cal D}+im)G_L(x,y)
(\stackrel{\leftarrow}{\not\!\!{\cal D}}-im)\right\} \nonumber \\ 
& &=\ \frac{1}{6}\left\{tr\ \gamma^5\not\!\!{\cal D}G_R(x,y)
\stackrel{\leftarrow}{\not\!\!{\cal D}}+tr\ \gamma^5\not\!\!{\cal D}G_L(x,y)
\stackrel{\leftarrow}{\not\!\!{\cal D}}\right. \nonumber \\
& &\hspace*{2cm}\left.+m^2 tr\ \gamma^5 G_R(x,y)+m^2 tr\ \gamma^5 G_L(x,y)\right\}_{g=1}\ \ .\label{4.9} 
\end{eqnarray}
The lower limit $g=0$ does not contribute as it involves the free Green's functions.
Eq.(\ref{4.9}), which is in five-dimensions because of the parameter $t$, should be contrasted
with the usual equation for the anomaly in 4-dimensions as given by Fujikawa's regularization [9]:
\begin{eqnarray}
tr\ \gamma^5\sum_n\varphi_n(x)\varphi_n^{\dag}(y)|_{reg}&=&\lim_{m\rightarrow\infty}tr\ \gamma^5
e^{-\frac{\not{\cal D}^2}{m^2}}\sum_n\varphi_n(x)\varphi_n^{\dag}(y) \nonumber \\
&=&\lim_{m\rightarrow\infty}tr\ \gamma^5e^{-\frac{\not{\cal D}^2}{m^2}}\int\frac{d^4k}{(2\pi)^4}e^{ik\cdot(x-y)}
\ .
\label{4.10}
\end{eqnarray}
We now outline how the terms on the RHS of (\ref{4.9}) are calculated. Using
$\not\!\!{\cal D}^2={\cal D}^2+\frac{1}{2}\gamma^\rho\gamma^\sigma G_{\rho\sigma}(x)$ $({\cal D}_\rho=\partial_\rho
+{\cal A}_\rho)$, the proper-time representation of the Green's function $G_R(x,y)$ leads to
\begin{eqnarray}
G_R(x,y)&=& \int\frac{d^4 k}{(2\pi)^4} e^{i k\cdot(x-y)}\int_0^\infty d\tau\  e^{-\tau ({\cal D}^2+2ik\cdot {\cal D}-
k^2+\frac{1}{2}\gamma^\rho\gamma^\sigma G_{\rho\sigma}+m^2)} \nonumber \\
&=&\int\frac{d^4 k}{(2\pi)^4} e^{i k\cdot(x-y)}\int_0^\infty d\tau\  e^{-\tau(m^2-k^2)}\sum_{n=0}^\infty
\frac{(-1)^n\ \tau^n}{n!}\left({\cal D}^2+2i k\cdot {\cal D}+\frac{1}{2}\gamma^\rho\gamma^\sigma G_{\rho\sigma}(x)
\right)^n \nonumber \\
&=& \int\frac{d^4 k}{(2\pi)^4} e^{i k\cdot(x-y)}\frac{1}{m^2-k^2}\sum_{n=0}^\infty (-1)^n \left(
\frac{{\cal D}^2+2ik\cdot {\cal D} %
+\frac{1}{2}\gamma^\rho\gamma^\sigma G_{\rho\sigma}(x)}{m^2-k^2}\right)^n \ .
\label{4.11}
\end{eqnarray}
Eq. (\ref{4.11}) shows that $(\not\!\! {\cal D}^2+m^2)G_R(x,y)=\delta (x-y)$, as we expect [18].
We then obtain
\begin{eqnarray}
\lefteqn{tr \ \gamma^5\not\!\!{\cal D}G_R(x,y)\stackrel{\leftarrow}{\not\!\!{\cal D}}}   \nonumber \\
& &=\ \ \ -tr\ \gamma^5 \gamma^\mu\gamma^\lambda\int\frac{d^4 k}{(2\pi)^4} 
e^{ik\cdot(x-y)}({\cal D}_\lambda + ik_\lambda) \nonumber \\
& &\ \ \ \ \times  \frac{1}{m^2-k^2}\sum_{n=0}^\infty (-1)^n \left(\frac{{\cal D}^2+2ik\cdot {\cal D}%
+\frac{1}{2}\gamma^\rho\gamma^\sigma
G_{\rho\sigma}(x)}{m^2-k^2}\right)^n(ik_\mu +{\cal A}_{\mu}(y))
\label{4.12}
\end{eqnarray}
Using
\[\gamma^\mu\gamma^\lambda=g^{\mu\lambda}+\sigma^{\mu\lambda},\ \ \   
\left(g^{\mu\lambda}=\frac{1}{2}\left\{\gamma^\mu,\gamma^\lambda\right\}, \ 
\sigma^{\mu\lambda}=\frac{1}{2}\left[\gamma^\mu,\gamma^\lambda\right]\right),
\]
and
\[tr\ \gamma^5\gamma^\lambda\gamma^\mu\gamma^\rho\gamma^\sigma=-4\epsilon^{\lambda\mu\rho\sigma}\ \ \  
 (\gamma^5=\gamma^4\gamma^1\gamma^2\gamma^3\ ,\epsilon^{1234}=+1),\]
 we find
\begin{eqnarray}
\lefteqn{tr \ \gamma^5\not\!\!{\cal D}G_R(x,y)\stackrel{\leftarrow}{\not\!\!{\cal D}}} \nonumber   \\
 & &=\ \ \  \int\frac{d^4 k}{(2\pi)^4} e^{i k\cdot(x-y)}({\cal D}^\nu + ik^\nu)\epsilon^{\lambda\mu\rho\sigma}
\frac{G_{\lambda\mu}(x)G_{\rho\sigma}(x)}{(m^2-k^2)^3}(ik_\nu +{\cal A}_{\nu}(y)) \nonumber   \\
  & &\ \ \ \ +2\int\frac{d^4 k}{(2\pi)^4} e^{i k\cdot(x-y)}({\cal D}_\lambda + ik_\lambda)\epsilon^{\mu\lambda\rho\sigma}
 \nonumber \\
 & &\ \ \ \ \times \left\{- \frac{1}{(m^2-k^2)^2}G_{\rho\sigma}(x)
 +\frac{1}{(m^2-k^2)^3}\left[2i k\cdot {\cal D}G_{\rho\sigma}(x)
+G_{\rho\sigma}(x)2i k\cdot {\cal D}\right]\right\}(ik_\mu+{\cal A}_{\mu}(y)). 
\label{4.13}
\end{eqnarray}
The proper-time representation of $G_L(x,y)$ using  
 ${\stackrel{\leftarrow}{\not\!\!{\cal D}^2}=\stackrel{\leftarrow}{{\cal D}^2}+\frac{1}{2}\gamma^\rho\gamma^\sigma 
G_{\rho\sigma}(y)}$ $\left(\stackrel{\leftarrow}{{\cal D}}_\rho=-\stackrel{\leftarrow}{\partial}_\rho
+{\cal A}_\rho(y)\right)$ is
\begin{eqnarray}
G_L(x,y)&=&\int\frac{d^4 k}{(2\pi)^4}\int_0^\infty d\tau\ e^{-\tau (\stackrel{\leftarrow}{{\cal D}^2}+2ik\cdot 
\stackrel{\leftarrow}{{\cal D}}-k^2+\frac{1}{2}\gamma^\rho\gamma^\sigma 
G_{\rho\sigma}(y)+m^2)}e^{i k\cdot(x-y)} \nonumber \\
&=&\int\frac{d^4 k}{(2\pi)^4}\frac{1}{m^2-k^2}\sum_{n=0}^\infty (-1)^n 
\left(\frac{\stackrel{\leftarrow}{{\cal D}^2}+2ik\cdot \stackrel{\leftarrow}{{\cal D}}%
+\frac{1}{2}\gamma^\rho\gamma^\sigma G_{\rho\sigma}(y)}{m^2-k^2}\right)^n e^{i k\cdot(x-y)} \label{4.14}
\end{eqnarray}
This gives
\begin{eqnarray}
\lefteqn{tr \ \gamma^5\not\!\!{\cal D}G_L(x,y)\stackrel{\leftarrow}{\not\!\!{\cal D}}} \nonumber   \\
& &=\    \int\frac{d^4 k}{(2\pi)^4}(ik^\nu +{\cal A}^{\nu})\epsilon^{\lambda\mu\rho\sigma}
\frac{1}{(m^2-k^2)^3}G_{\lambda\mu}(y)G_{\rho\sigma}(y)(\stackrel{\leftarrow}{{\cal D}}_\nu + ik_\nu)
e^{ik\cdot(x-y)} \nonumber   \\
& &\ \ +2\int\frac{d^4 k}{(2\pi)^4}(ik_\lambda+{\cal A}_{\lambda}) \epsilon^{\mu\lambda\rho\sigma} \nonumber \\
& &\ \ \times \left\{- \frac{1}{(m^2-k^2)^2}G_{\rho\sigma}(y)+\frac{1}{(m^2-k^2)^3}
\left[2i k\cdot \stackrel{\leftarrow}{{\cal D}}G_{\rho\sigma}(y)
+G_{\rho\sigma}(y)2ik\cdot \stackrel{\leftarrow}{{\cal D}}\right]\right\}
(\stackrel{\leftarrow}{{\cal D}}_\mu +ik_\mu)e^{i k\cdot(x-y)}.
\label{4.15}
\end{eqnarray}
From (\ref{4.11}) and (\ref{4.14}), we also obtain
\begin{equation}
m^2tr\ \gamma^5G_R(x,y)=- m^2\int\frac{d^4 k}{(2\pi)^4}e^{i k\cdot(x-y)}\epsilon^{\lambda\mu\rho\sigma}
\frac{1}{(m^2-k^2)^3}G_{\lambda\mu}(x)G_{\rho\sigma}(x)\ ,
\label{4.16}
\end{equation}
\begin{equation}
m^2tr\ \gamma^5G_L(x,y)=- m^2\int\frac{d^4 k}{(2\pi)^4}e^{i k\cdot(x-y)}\epsilon^{\lambda\mu\rho\sigma}
\frac{1}{(m^2-k^2)^3}G_{\lambda\mu}(y)G_{\rho\sigma}(y)\ .
\label{4.17}
\end{equation}
At this point, we simplify Eqs. (\ref{4.13}) and (\ref{4.15}) by dropping terms of $O(\frac{1}{m})$, as we are
interested in the limit $m\rightarrow\infty$:
\begin{eqnarray}
\lefteqn{tr \ \gamma^5\not\!\!{\cal D}G_R(x,y)\stackrel{\leftarrow}{\not\!\!{\cal D}}} \nonumber   \\
& &\simeq\   -\int\frac{d^4 k}{(2\pi)^4}e^{i k\cdot(x-y)}\epsilon^{\lambda\mu\rho\sigma}
\frac{k^2}{(m^2-k^2)^3}G_{\lambda\mu}(x)G_{\rho\sigma}(x)\nonumber \\
& &\ \ -2\int\frac{d^4 k}{(2\pi)^4}e^{i k\cdot(x-y)}\epsilon^{\mu\lambda\rho\sigma}
\frac{1}{(m^2-k^2)^2}\left\{{\cal D}_\lambda G_{\rho\sigma}(x)ik_\mu+{\cal D}_\lambda G_{\rho\sigma}(x){\cal A}_{\mu}(y)
+ik_\lambda G_{\rho\sigma}(x){\cal A}_{\mu}(y)\right\} \nonumber \\
& &\ \ +2\int\frac{d^4 k}{(2\pi)^4}e^{i k\cdot(x-y)}\epsilon^{\mu\lambda\rho\sigma}\frac{1}{(m^2-k^2)^3}
 \left\{{\cal D}_\lambda\left[2i k\cdot {\cal D}G_{\rho\sigma}(x)
+ G_{\rho\sigma}(x)2i k\cdot{\cal D}\right]ik_\mu\right. \nonumber \\
& &\hspace*{7.5cm}\left.+ ik_\lambda\left[2ik\cdot {\cal D}G_{\rho\sigma}(x)+G_{\rho\sigma}(x)2i k\cdot{\cal D}\right]
{\cal A}_{\mu}(y)\right\}.\label{4.18}
\end{eqnarray}
\begin{eqnarray}
\lefteqn{tr \ \gamma^5\not\!\!{\cal D}G_L(x,y)\stackrel{\leftarrow}{\not\!\!{\cal D}}} \nonumber   \\
& &\simeq\  -\int\frac{d^4 k}{(2\pi)^4}e^{i k\cdot(x-y)}\epsilon^{\lambda\mu\rho\sigma}
\frac{k^2}{(m^2-k^2)^3}G_{\lambda\mu}(y)G_{\rho\sigma}(y)\nonumber \\
& &\ \ \  -2\int\frac{d^4 k}{(2\pi)^4}e^{i k\cdot(x-y)}\epsilon^{\mu\lambda\rho\sigma}\frac{1}{(m^2-k^2)^2}\nonumber \\
& & \hspace*{3cm}\times\left\{ik_\lambda G_{\rho\sigma}(y)\stackrel{\leftarrow}{{\cal D}}_\mu+
{\cal A}_\lambda(x) G_{\rho\sigma}(y)\stackrel{\leftarrow}{{\cal D}}_\mu+{\cal A}_\lambda(x) G_{\rho\sigma}(y)ik_\mu \right\}
\nonumber \\
& &\ \ \  +2\int\frac{d^4 k}{(2\pi)^4}e^{ik\cdot(x-y)}\epsilon^{\mu\lambda\rho\sigma}\frac{1}{(m^2-k^2)^3}
\left\{ik_\lambda\left[2ik\cdot\stackrel{\leftarrow}{{\cal D}}G_{\rho\sigma}(y)+
G_{\rho\sigma}(y)2i k\cdot\stackrel{\leftarrow}{{\cal D}}\right]\stackrel{\leftarrow}{{\cal D}}_\mu\right.\nonumber \\
& & \hspace*{7.5cm}\left.+{\cal A}_\lambda(x)\left[2ik\cdot\stackrel{\leftarrow}{{\cal D}}G_{\rho\sigma}(y)+
G_{\rho\sigma}(y)2ik\cdot\stackrel{\leftarrow}{{\cal D}}\right]ik_\mu\right\}.
\nonumber \\
\label{4.19}
\end{eqnarray}
(Terms having $k_\lambda k_\mu$ in their integrand are dropped, since they vanish due to the
anti- symmetric tensor $\epsilon^{\lambda\mu\rho\sigma}.)$

Let us consider the coincidence limit $(y\rightarrow x)$ of 
$\displaystyle{tr\ \gamma^5\sum_n\varphi_n(x,t)\varphi_n^{\dag}(y,t)|_{reg}}$ given by Eq.(4.9).
From Eq.(\ref{4.18}), we find
\begin{eqnarray}
\lefteqn{\lim_{y \rightarrow x}tr \ \gamma^5\not\!\!{\cal D}G_R(x,y)\stackrel{\leftarrow}{\not\!\!{\cal D}}} \nonumber \\  
& &=\ -\epsilon^{\lambda\mu\rho\sigma}\int\frac{d^4k}{(2\pi)^4}  
\frac{k^2}{(m^2-k^2)^3}G_{\lambda\mu}(x)G_{\rho\sigma}(x)\nonumber \\
& &\ \ \ -2\epsilon^{\mu\lambda\rho\sigma}\int\frac{d^4k}{(2\pi)^4}
\frac{1}{(m^2-k^2)^2}\left\{{\cal D}_\lambda G_{\rho\sigma}(x)ik_\mu+{\cal D}_\lambda 
G_{\rho\sigma}(x){\cal A}_{\mu}(y)
+ik_\lambda G_{\rho\sigma}(x){\cal A}_{\mu}(y)\right\}_{y=x} \nonumber \\
& &\ \ \ +2\epsilon^{\mu\lambda\rho\sigma}\int\frac{d^4k}{(2\pi)^4}\frac{1}{(m^2-k^2)^3}
\left\{{\cal D}_\lambda\left[2i k\cdot {\cal D}G_{\rho\sigma}(x)
+G_{\rho\sigma}(x)2i k\cdot{\cal D}\right]ik_\mu\right. \nonumber \\
& &\hspace*{6.0cm}\left.+ik_\lambda\left[2ik\cdot{\cal D}G_{\rho\sigma}(x)+G_{\rho\sigma}(x)2i k\cdot{\cal %
D}\right]{\cal A}_{\mu}(y) \right\}_{y=x}\ .
\label{4.20}
\end{eqnarray}
Evaluation of the RHS of Eq. (\ref{4.20}) is considerably simplified, if we note
\[ 
G_{\rho\sigma}(x)=[{\cal D}_\rho,{\cal D}_\sigma],\ \ \epsilon^{\lambda\mu\rho\sigma}G_{\rho\sigma}(x)=
2\epsilon^{\lambda\mu\rho\sigma}{\cal D}_\rho{\cal D}_\sigma,
\]
\[
2\epsilon^{\lambda\mu\rho\sigma}{\cal D}_\lambda G_{\rho\sigma}{\cal D}_\mu=4\epsilon^{\lambda\mu\rho\sigma}
{\cal D}_\lambda{\cal D}_\rho{\cal D}_\sigma{\cal D}_\mu=\epsilon^{\lambda\mu\rho\sigma}G_{\lambda\mu}
G_{\rho\sigma}\ \ ;
\]
furthermore, 
\[
\int d^4k\ k_\mu\ f(k^2)=0\ ,\ 
\int d^4k\ k_\lambda k_\mu f(k^2)=\delta_{\lambda\mu}\int d^4k \frac{1}{4}(-k^2)f(k^2).
\]
The divergent terms in Eq.(\ref{4.20}) cancel, and the equation becomes
\begin{eqnarray}
\lim_{y \rightarrow x}tr\ \gamma^5\not\!\!{\cal D}G_R(x,y)\stackrel{\leftarrow}{\not\!\!{\cal D}}
& = & -2\epsilon^{\mu\lambda\rho\sigma}\int\frac{d^4k}{(2\pi)^4}\frac{m^2}{(m^2-k^2)^3}
{\cal D}_\lambda G_{\rho\sigma}(x){\cal A}_\mu (y)|_{y=x} \nonumber \\ 
 & = &  \frac{1}{16\pi^2}\epsilon^{\lambda\mu\rho\sigma}G_{\rho\sigma}(x){\cal A}_\lambda (x){\cal A}_\mu (x)\ \
 .
\label{4.21}
\end{eqnarray}
At the last step, we have used the Bianchi identity $\epsilon^{\mu\lambda\rho\sigma} D_\lambda G_{\rho\sigma}(x)=0$ 
to write $\epsilon^{\mu\lambda\rho\sigma}{\cal D}_\lambda G_{\rho\sigma}(x)=
\epsilon^{\mu\lambda\rho\sigma}G_{\rho\sigma}(x){\cal A}_\lambda (x)$.
($D_\lambda=\partial_\lambda +[{\cal A}_\lambda,\ ],{\cal D}_\lambda=\partial_\lambda+{\cal A}_\lambda$.)
From Eq.(\ref{4.19}), we have
\begin{eqnarray}
\lefteqn{\lim_{y\rightarrow x}tr \ \gamma^5\not\!\!{\cal D}G_L(x,y)\stackrel{\leftarrow}{\not\!\!{\cal D}}} \nonumber \\
&=& -\epsilon^{\lambda\mu\rho\sigma}\int\frac{d^4k}{(2\pi)^4}\frac{k^2}{(m^2-k^2)^3}
G_{\lambda\mu}(y)G_{\rho\sigma}(y)\nonumber \\
& & -2\epsilon^{\mu\lambda\rho\sigma}\int\frac{d^4k}{(2\pi)^4}\frac{1}{(m^2-k^2)^2}
\left\{ik_\lambda G_{\rho\sigma}(y)\stackrel{\leftarrow}{{\cal D}}_\mu+
{\cal A}_\lambda(x) G_{\rho\sigma}(y)\stackrel{\leftarrow}{{\cal D}}_\mu\right. \nonumber \\
& & \hspace*{8cm}\left.+{\cal A}_\lambda(x) G_{\rho\sigma}(y)\stackrel{ }{ik_\mu}\right\}_{y=x} \nonumber \\
& & +2\epsilon^{\mu\lambda\rho\sigma}\int\frac{d^4k}{(2\pi)^4}\frac{1}{(m^2-k^2)^3}
\left\{ik_\lambda\left[2i k\cdot\stackrel{\leftarrow}{{\cal D}}G_{\rho\sigma}(y)+
G_{\rho\sigma}(y)2ik\cdot\stackrel{\leftarrow}{{\cal D}}\right]\stackrel{\leftarrow}{{\cal D}_\mu}\right.\nonumber \\
& & \hspace*{6.0cm}\left.+{\cal A}_\lambda(x)\left[2ik\cdot\stackrel{\leftarrow}{{\cal D}}G_{\rho\sigma}(y)+
G_{\rho\sigma}(y)2ik\cdot\stackrel{\leftarrow}{{\cal D}}\right]ik_\mu\right\}_{y=x}
\label{4.22}
\end{eqnarray}
To evaluate the RHS of (\ref{4.22}), we use:
\[
G_{\rho\sigma}(y)=[\stackrel{\leftarrow}{{\cal D}}_\rho,\stackrel{\leftarrow}{{\cal D}}_\sigma], \ \ 
\epsilon^{\lambda\mu\rho\sigma}G_{\rho\sigma}(y)=2\epsilon^{\lambda\mu\rho\sigma}\stackrel{\leftarrow}{{\cal D}}_\rho
\stackrel{\leftarrow}{{\cal D}}_\sigma,
\]
\[
 2\epsilon^{\lambda\mu\rho\sigma}\stackrel{\leftarrow}{{\cal D}}_\lambda G_{\rho\sigma}(y)
 \stackrel{\leftarrow}{{\cal D}}_\mu
 =4\epsilon^{\lambda\mu\rho\sigma}
 \stackrel{\leftarrow}{{\cal D}}_\lambda\stackrel{\leftarrow}{{\cal D}}_\rho
 \stackrel{\leftarrow}{{\cal D}}_\sigma\stackrel{\leftarrow}{{\cal D}}_\mu=
 \epsilon^{\lambda\mu\rho\sigma}G_{\lambda\mu}(y) G_{\rho\sigma}(y)\ .
 \]
 As before, the divergent terms in (\ref{4.22}) cancel, and we arrive  at
 \begin{eqnarray}
 \lim_{y \rightarrow x}tr\ \gamma^5\not\!\!{\cal D}G_L(x,y)\stackrel{\leftarrow}{\not\!\!{\cal D}}
 & = & -2\epsilon^{\mu\lambda\rho\sigma}\int\frac{d^4k}{(2\pi)^4}\frac{m^2}{(m^2-k^2)^3}
 {\cal A}_\lambda (x)G_{\rho\sigma}(y)\stackrel{\leftarrow}{{\cal D}}_\mu|_{y=x} \nonumber \\
 &=&\frac{1}{16\pi^2}\epsilon^{\lambda\mu\rho\sigma}{\cal A}_\lambda (x){\cal A}_\mu (x)G_{\rho\sigma}(x)
 \label{4.23}
 \end{eqnarray}
 From Eqs.(\ref{4.16}) and (\ref{4.17}), we get
\begin{equation}
\lim_{y \rightarrow x}m^2tr\ \gamma^5G_R(x,y)=-\frac{1}{32\pi^2}\epsilon^{\lambda\mu\rho\sigma}G_{\lambda\mu}(x)%
G_{\rho\sigma}(x)\ \ ,
\label{4.24}
\end{equation}
\begin{equation}
\lim_{y \rightarrow x}m^2tr\ \gamma^5G_L(x,y)=-\frac{1}{32\pi^2}\epsilon^{\lambda\mu\rho\sigma}G_{\lambda\mu}(x)%
G_{\rho\sigma}(x)\ .
\label{4.25}
\end{equation}
Inserting the above results in (\ref{4.9}), we obtain
\begin{eqnarray}
\lefteqn{\lim_{y\rightarrow x}tr\ \gamma^5\sum_n\varphi_n(x,t)\varphi_n^{\dag}(y,t)|_{reg}} \nonumber \\
& &\ \ = -\frac{1}{24\pi^2} \epsilon^{\lambda\mu\rho\sigma}\partial_\lambda\left({\cal A}_\mu\partial_\rho
{\cal A}_\sigma + \frac{1}{2} {\cal A}_\mu{\cal A}_\rho{\cal A}_\sigma\right)
-\frac{1}{48\pi^2}\epsilon^{\lambda\mu\rho\sigma}{\cal A}_\lambda\partial_\mu{\cal A}_\rho{\cal A}_\sigma
\label{4.26}
\end{eqnarray}
The first term on the RHS of (\ref{4.26}),
in fact, yields the consistent anomaly. The presence of
the extra term on the RHS of (\ref{4.26}) shows that a simple
coincidence limit of ${\displaystyle tr\ \gamma^5\sum_n %
\varphi_n(x,t)\varphi_n^{\dag}(y,t)|_{reg}}$ does not satisfy
the consistency condition. In the next section, we show that if
the point-splitting of 
${\displaystyle tr\ \gamma^5\sum_n\varphi_n(x,t)\varphi_n %
^{\dag}(x,t)}$
is done in a gauge invariant way, then besides the terms on the
RHS of (\ref{4.26}), additional terms appear. These terms are formally of 
$O(\epsilon^2)$, where $\epsilon = x-y$. In the coincidence limit,
they provide a finite contribution that cancel the last term
in (\ref{4.26}) and yield the consistent anomaly. Clearly, 
a gauge invariant point-splitting regularization is needed
in the evaluation of the anomaly. This agrees with the Abelian
case, where the gauge invariant point-splitting leads to the 
$U(1)_A$ anomaly [19-21].
\section{Anomaly Evaluation by Gauge Invariant Point-Splitting }
We examine now the evaluation of the anomaly as given by the gauge invariant point-splitting
technique. To this end, we note that the wavefunction $\varphi_n^{\dag}(x)$ can be related to the
wavefunction $\varphi_n^{\dag}(y)$ by a path-ordered gauge transformation:
\[
\varphi_n^{\dag}(x)=\varphi_n^{\dag}(y)\Omega(y,x)\ ,
\]
where $\displaystyle{\Omega(y,x)=P\left[e^{-\int_x^y A^\mu(x^{\prime})dx^{\prime}_\mu}\right]}$ [22]. 
Therefore, we can write
\begin{equation}
tr\ \gamma^5\sum_n\varphi_n(x)\varphi^{\dag}_n(x)=tr\ \gamma^5\sum_n\varphi_n(x)\varphi^{\dag}_n(y)\Omega(y,x)\
.
\label{5.1}
\end{equation}
Differentiating this with respect to $y$, we get
\[
0= \frac{\partial}{\partial y^\alpha}tr\left[\ \gamma^5\sum_n\varphi_n(x)\varphi^{\dag}_n(y)\right]\cdot\Omega(y,x)
-tr\left[\ \gamma^5\sum_n\varphi_n(x)\varphi^{\dag}_n(y)\right]{\cal A}_\alpha(y)\Omega(y,x),
\]
and hence,
\begin{equation}
tr\left[\ \gamma^5\sum_n\varphi_n(x)\varphi^{\dag}_n(y)\right]{\cal A}_\alpha(y)=
\frac{\partial}{\partial y^\alpha}tr\left[\ \gamma^5\sum_n\varphi_n(x)\varphi^{\dag}_n(y)\right]\ \ .
\label{5.2}
\end{equation}
We next consider $y$ to be infinitesimally separate from $x$, i.e., $x-y=\epsilon$, where $\epsilon\rightarrow 0$. Then, from 
Eqs. (\ref{5.1}) and (\ref{5.2}),
\begin{eqnarray}
\lefteqn{tr\ \gamma^5\sum_n\varphi_n(x)\varphi^{\dag}_n(x)} \nonumber \\
& & =  \lim_{y\rightarrow x}\left[ tr\ \gamma^5\sum_n\varphi_n(x)\varphi^{\dag}_n(y)-tr \
\gamma^5\sum_n\varphi_n(x)\varphi^{\dag}_n(y){\cal A}_\alpha(y)(y-x)^\alpha\right] \nonumber \\
& & =  \lim_{y\rightarrow x}\left[ tr\ \gamma^5\sum_n\varphi_n(x)\varphi^{\dag}_n(y)
+(x-y)^\alpha \frac{\partial}{\partial y^{\alpha}}tr\ \gamma^5\sum_n\varphi_n(x)\varphi^{\dag}_n(y)\right]\ .
\label{5.3}
\end{eqnarray}
As before, in (\ref{5.3}) the dependence on the parameter $t$ is suppressed. In the previous section, we
obtained a regularized expression for the first term on the RHS of (\ref{5.3}), i.e., for
$\displaystyle{tr\ \gamma^5\sum_n\varphi_n(x)\varphi^{\dag}_n(y)}$ given by Eq.(\ref{4.9}) 
and Eqs.(\ref{4.16}-\ref{4.19}). Also, we pointed out
that a simple coincidence limit of the expression does not lead to the consistent anomaly. This 
suggests that the second term in (\ref{5.3}) may produce a finite contribution in the coincidence limit, which
combined with the contribution of the first term, may lead to the consistent anomaly. We now investigate this point.

Eq.(\ref{4.9}) and Eqs.(\ref{4.16}-\ref{4.19}) show that $\displaystyle{tr\ \gamma^5\sum_n\varphi_n(x,t)\varphi^{\dag}_n(y,t)
|_{reg}}$ is of the following form:
\begin{equation}
tr\ \gamma^5\sum_n\varphi_n(x,t)\varphi^{\dag}_n(y,t)|_{reg}=
\int\frac{d^4k}{(2\pi)^4}e^{i k\cdot(x-y)}\left[f(k;x)+g(k;y)+h(k;x,y)\right]\ ,
\label{5.4}
\end{equation}
where the functions $f(k;x),\ g(k;y), \mbox{ and }h(k;x,y)$ can be read off from the RHS of (\ref{4.16}-\ref{4.19}).
Let us examine the contribution of each of these terms to the RHS of (\ref{5.3}):
\begin{eqnarray}
\lefteqn{\int\frac{d^4k}{(2\pi)^4}e^{ik\cdot(x-y)}f(k;x)+(x-y)^{\alpha}\frac{\partial}%
{\partial y^\alpha}\int\frac{d^4k}{(2\pi)^4}e^{i k\cdot(x-y)}f(k;x)} \nonumber \\
& & = \ \int\frac{d^4k}{(2\pi)^4}e^{i k\cdot(x-y)}f(k;x)+
\int\frac{d^4k}{(2\pi)^4}e^{ik\cdot(x-y)}\frac{\partial}{\partial k_\alpha}
\left[k_\alpha f(k;x)\right] \nonumber \\ 
& & =\ \int\frac{d^4k}{(2\pi)^4}f(k;x)+\int\frac{d^4k}{(2\pi)^4}\frac{\partial}{\partial k_\alpha}
\left[k_\alpha f(k;x)\right]\ , \label{5.5}
\end{eqnarray}
in the coincidence limit $(y\rightarrow x)$.
\begin{eqnarray}
\lefteqn{\int\frac{d^4k}{(2\pi)^4}e^{ik\cdot(x-y)}g(k;y)+(x-y)^{\alpha}\frac{\partial}%
{\partial y^\alpha}\int\frac{d^4k}{(2\pi)^4}e^{i k\cdot(x-y)}g(k;y)} \nonumber \\
& & = \ \int\frac{d^4k}{(2\pi)^4}e^{i k\cdot(x-y)}g(k;x)+
\int\frac{d^4k}{(2\pi)^4}e^{ik\cdot(x-y)}\frac{\partial}{\partial k_\alpha} 
\left[k_\alpha g(k;y)\right] \nonumber \\
& & =\ \int\frac{d^4k}{(2\pi)^4}g(k;x)+\int\frac{d^4k}{(2\pi)^4}\frac{\partial}{\partial k_\alpha}
\left[k_\alpha g(k;x)\right]\ , \label{5.6}
\end{eqnarray}
again in the coincidence limit.
The integrals over $h(k;x,y)$ need more care:
\begin{eqnarray}
\lefteqn{\int\frac{d^4k}{(2\pi)^4}e^{ik\cdot(x-y)}h(k;x,y)+(x-y)^{\alpha}\frac{\partial}%
{\partial y^\alpha}\int\frac{d^4k}{(2\pi)^4}e^{ik\cdot(x-y)}h(k;x,y)} \nonumber \\
& &= \ \ \int\frac{d^4k}{(2\pi)^4}e^{ik\cdot(x-y)}h(k;x,x)+(x-y)^{\alpha}
\int\frac{d^4k}{(2\pi)^4}e^{ik\cdot(x-y)}(-ik_\alpha)h(k;x,y) \nonumber \\
& &= \ \ \int\frac{d^4k}{(2\pi)^4}e^{ik\cdot(x-y)}h(k;x,x)+(x-y)^{\alpha}
\int\frac{d^4k}{(2\pi)^4}e^{ik\cdot(x-y)}(-ik_\alpha) \nonumber \\
& & \hspace*{7.5cm}\times\left[h(k;x,x)-(x-y)^{\beta}\frac{\partial}{\partial y^\beta}h(k;x,y)\right] \nonumber \\
& &= \ \ \int\frac{d^4k}{(2\pi)^4}e^{ik\cdot(x-y)}h(k;x,x)+
\int\frac{d^4k}{(2\pi)^4}e^{ik\cdot(x-y)}\frac{\partial}{\partial k_\alpha}
\left[k_\alpha h(k;x,x)\right] \nonumber \\
& &\hspace*{3.0cm}+(x-y)^{\alpha}(x-y)^{\beta}\int\frac{d^4k}{(2\pi)^4}e^{ik\cdot(x-y)}
ik_\alpha\frac{\partial}{\partial y^\beta}h(k;x,y) \nonumber \\
& & =\ \ \int\frac{d^4k}{(2\pi)^4}h(k;x,x)
+\int\frac{d^4k}{(2\pi)^4}\frac{\partial}{\partial k_\alpha}
\left[k_\alpha h(k;x,x)\right] \nonumber \\
& &\hspace*{3cm}-5i\int\frac{d^4k}{(2\pi)^4}\frac{\partial}{\partial k_\alpha}
\frac{\partial}{\partial y^\alpha}h(k;x,y)|_{y=x} \nonumber \\
& &\hspace*{4cm}-i\int\frac{d^4k}{(2\pi)^4}k_\alpha\frac{\partial^2}{\partial k_\alpha %
\partial k_\beta}\frac{\partial}{\partial y^\beta}h(k;x,y)|_{y=x}\ ,
\label{5.7}
\end{eqnarray}
in the coincidence limit.
We note that the second term on the RHS of (\ref{5.7}) is formally of $O(\epsilon)$,
whereas the last two terms are formally of $O(\epsilon^2)$.

Let us give an example to show how a finite contribution can emerge from the last two terms in
(\ref{5.7}). This also illustrates how we carry out these calculations. We consider $h(k;x,y)$ having
a term like
\[
\tilde{h}(k;x,y)=i\epsilon^{\mu\lambda\rho\sigma}\frac{k_\mu\xi_{\lambda\rho\sigma}(x,y)}{(m^2-k^2)^2}\ ,
\]
which is odd in $k$ (in fact, such terms occur in $h(k;x,y)$ as seen from Eqs.(\ref{4.18}) and (\ref{4.19})).
Here, $\xi_{\lambda\rho\sigma}(x,y)$ is some polynomial in the gauge field and its derivative. Inserting
$\tilde{h}(k;x,y)$ for $h(k;x,y)$ in the last two terms in (\ref{5.7}), we obtain
\begin{eqnarray}
\lefteqn{ 5\int\!\frac{d^4k}{(2\pi)^4}\epsilon^{\mu\lambda\rho\sigma}\left\{\frac{\delta_{\mu\alpha}}{(m^2-k^2)^2}
-\frac{4k_\mu k_\alpha}{(m^2-k^2)^3}\right\}\frac{\partial}{\partial y^\alpha}\xi_{\lambda\rho\sigma}(x,y)|_{y=x}}
\nonumber \\
& & \hspace*{1cm}+\int\!\frac{d^4k}{(2\pi)^4}\epsilon^{\mu\lambda\rho\sigma}k_\alpha\frac{\partial}{\partial k_\alpha}
\left\{\frac{\delta_{\mu\beta}}{(m^2-k^2)^2}
-\frac{4k_\mu k_\beta}{(m^2-k^2)^3}\right\}\frac{\partial}{\partial y^\beta}\xi_{\lambda\rho\sigma}(x,y)|_{y=x}
\nonumber \\
& & = \ \ 5\int\!\frac{d^4k}{(2\pi)^4}\epsilon^{\mu\lambda\rho\sigma}\frac{m^2}{(m^2-k^2)^3}
\frac{\partial \xi_{\lambda\rho\sigma}(x,y)}{\partial y^\mu}|_{y=x} \nonumber \\
& &\hspace*{1cm} +6\int\!\frac{d^4k}{(2\pi)^4}\epsilon^{\mu\lambda\rho\sigma}\frac{m^2\ k^2}{(m^2-k^2)^4}
\frac{\partial \xi_{\lambda\rho\sigma}(x,y)}{\partial y^\mu}|_{y=x} \nonumber \\
& & = \frac{1}{32\pi^2}\epsilon^{\mu\lambda\rho\sigma}\frac{\partial \xi_{\lambda\rho\sigma}(x,y)}{\partial %
y^\mu}|_{y=x}\ , \label{5.8}
\end{eqnarray}
which is a finite contribution. We also note that, because $\tilde{h}(k;x,y)$ is odd in $k$, its
contribution to the first two terms on the RHS of (\ref{5.7}) is zero.

From Eqs. (\ref{5.3}-\ref{5.7}), we now have 
\begin{eqnarray}
\lefteqn{tr \ \gamma^5\sum_n\varphi_n(x)\varphi_n^{\dag}(x)|_{reg}} \nonumber \\
& & \simeq \int\!\frac{d^4k}{(2\pi)^4}\left[f(k;x)+g(k;x)+h(k;x,x)\right] \nonumber \\ 
& & \hspace*{1cm}+\int\!\frac{d^4k}{(2\pi)^4}\frac{\partial}{\partial k_\alpha} \ 
\left(k_\alpha  \left[f(k;x)+g(k;x)+h(k;x,x)\right]\right)\nonumber \\
& & \hspace{2cm}-5i\int\frac{d^4k}{(2\pi)^4}\frac{\partial}{\partial k_\alpha}
\frac{\partial}{\partial y^\alpha}h(k;x,y)|_{y=x} \nonumber \\
& & \hspace{3cm}-i\int\frac{d^4k}{(2\pi)^4}k_\alpha\frac{\partial^2}{\partial k_\alpha %
\partial k_\beta}\frac{\partial}{\partial y^\beta}h(k;x,y)|_{y=x}\ .
\label{5.9}
\end{eqnarray}
At this point, we notice certain inherent freedom in the choice of $g(k;y)\mbox{ and } h(k;x,y)$.
If we change
\begin{equation}
g(k;y)\rightarrow g^\prime (k;y)=g(k;y)-\chi (k;y)
\label{5.10}
\end{equation}
and 
\begin{equation}
h(k;x,y)\rightarrow h^\prime (k;x,y)=h(k;x,y)+\chi (k;y)\ ,
\label{5.11}
\end{equation}
then the original expression for $\displaystyle{tr \ \gamma^5\sum_n\varphi_n(x)\varphi_n^{\dag}(y)|_{reg}}$
(Eq. (\ref{5.4})) remains the same. Now, on the RHS of Eq. (\ref{5.9}), the first two terms remain unchanged.
However, the last two terms (formally of order $\epsilon^2$) can change, since $h(k;x,y)$ is being replaced
by $h^\prime(k;x,y)$. This can result in a different expression for the anomaly.
The only restriction on the choice of $h^\prime(k;x,y)\mbox{ and } g^\prime(k;y)$ 
is that the anomaly has to satisfy
the WZ consistency condition. 

This restriction allows us only two choices.
One is to take $h^{\prime}(k;x,y)$ such that
in Eq.(\ref{4.19}) the following change occurs:
\begin{equation}
{\cal A}_\lambda(x) G_{\rho\sigma}(y)ik_\mu\rightarrow {\cal A}_\lambda(x) G_{\rho\sigma}(y)ik_\mu
+2{\cal A}_\lambda(y)\partial_\rho{\cal A}_\sigma (y) i k_\mu\ .
\label{5.12}
\end{equation}
Correspondingly, $g^\prime(k;y)$ has to be taken such that in the same equation
\begin{equation}
ik_{\lambda}G_{\rho\sigma}(y)\stackrel{\leftarrow}{{\cal D}_\mu}\rightarrow
ik_{\lambda}G_{\rho\sigma}(y)\stackrel{\leftarrow}{{\cal D}_\mu}+2ik_\lambda{\cal A}_\mu(y)\partial_\rho{\cal
A}_\sigma (y).
\label{5.13}
\end{equation}
With this $h^{\prime}(k;x,y)$, we find from the last two terms of (\ref{5.9}) the contribution:
\[
\frac{1}{48\pi^2} \epsilon^{\lambda\mu\rho\sigma}{\cal A}_\lambda(x)\partial_\mu{\cal A}_\rho(x)
{\cal A}_\sigma(x)\ .
\]
This contribution exactly cancels the extra term we obtain from the coincidence limit 
as given by Eq.(4.26). The contribution of the second term on the RHS of (\ref{5.9}) vanishes.
We, therefore, obtain
the consistent anomaly given by the differential
geometric approach as well as by the diagrammatic approach using Pauli-Villars regularization [23,24]:
\begin{eqnarray}
a({\cal A})&=&-\frac{1}{2} tr \ \gamma^5\sum_n\varphi_n(x,t)\varphi_n^{\dag}(x,t)|_{reg}\nonumber \\
&=& \frac{1}{48\pi^2}\epsilon^{\lambda\mu\rho\sigma}\partial_\lambda\left[{\cal A}_\mu
\partial_\rho{\cal A}_\sigma + \frac{1}{2} {\cal A}_\mu{\cal A}_\rho{\cal A}_\sigma \right]\ .
\label{5.14}
\end{eqnarray}
The other choice for $h^{\prime}(k;x,y)$ is to take it such that in Eq.(4.19)
\begin{equation}
{\cal A}_\lambda(x) G_{\rho\sigma}(y)ik_\mu\rightarrow{\cal A}_\lambda(x) G_{\rho\sigma}(y)ik_\mu+
ik_{\lambda}G_{\rho\sigma}(y)\stackrel{\leftarrow}{{\cal D}_\mu}\ ,
\label{5.15}
\end{equation}
and the corresponding $g^\prime(k;y)$ no longer contains the term
$ik_{\lambda}G_{\rho\sigma}(y)\stackrel{\leftarrow}{{\cal D}_\mu}$.
We find  from the last two terms in (\ref{5.9}) the contribution: 
\[
\frac{1}{24\pi^2} \epsilon^{\lambda\mu\rho\sigma}\partial_\lambda\left({\cal A}_\mu\partial_\rho
{\cal A}_\sigma + \frac{1}{2} {\cal A}_\mu{\cal A}_\rho{\cal A}_\sigma\right)
+\frac{1}{48\pi^2}\epsilon^{\lambda\mu\rho\sigma}{\cal A}_\lambda\partial_\mu{\cal A}_\rho{\cal A}_\sigma\ .
\]
This contribution exactly cancels the contribution of the first term  on the RHS of (\ref{5.9}) (as given by the
Eq.(\ref{4.26})). The contribution of the second term on the RHS of (\ref{5.9}) again vanishes. This leads to
\begin{eqnarray}
a({\cal A})&=& -\frac{1}{2} tr \ \gamma^5\sum_n\varphi_n(x,t)\varphi_n^{\dag}(x,t)|_{reg}\nonumber \\
&=& 0,
\label{5.16}
\end{eqnarray}
which, of course, is a perfectly acceptable result, since $D_\mu({\cal A})j^\mu(x,t)=0$ trivially
satisfies the WZ consistency condition. Clearly, the choice (\ref{5.12}) corresponds to an anomalous
theory with a nonvanishing WZW action, while the choice (\ref{5.15}) corresponds to a non-anomalous theory. 

The above two choices, therefore, show that either
\begin{equation}
D_\mu({\cal A})j^{\mu}_L(x,t)=\frac{1}{48\pi^2}\epsilon^{\lambda\mu\rho\sigma}\partial_\lambda\left[{\cal A}_\mu
\partial_\rho{\cal A}_\sigma + \frac{1}{2} {\cal A}_\mu{\cal A}_\rho{\cal A}_\sigma \right]\ ,
\label{5.17}
\end{equation}
or
\begin{equation}
D_\mu({\cal A})j^{\mu}_L(x,t)=0\ ,
\label{5.18}
\end{equation}
where the current $j^{\mu,a}_L(x,t)$ is the vacuum expectation value of an operator current (Eq.(2.27)):
\[
j^{\mu,a}_L(x,t)=\langle\Psi_o|\bar{\psi}_L(x)\gamma^\mu(-iT^a)\psi_L(x)|\Psi_o\rangle\ .
\]
${\cal A}_\mu(x,t)$ is the gauge transformed field
${\cal A}_\mu(x,t)= U^{-1}(x,t) A_\mu(x)U(x,t)+U^{-1}(x,t)\partial_\mu U(x,t)$ and $|\Psi_o\rangle$ is
the vacuum of the fully interacting system. The five dimensional
space $(x,t)$ in our case corresponds to a cylinder with its flat ends at $t=0$ and $t=1$. If we denote this
cylindrical space by $C^{5}_{1}$, then there is another cylindrical space $C^{5}_{2}$ with the same flat ends,
such that  $C^{5}_{1}-C^{5}_{2}=S^4\times S^1=S^5$, where $S^5$ is a five dimensional torus [13]. The
unitary field $U(x,t)$ provides a mapping of the five dimensional space $S^5$ onto the internal space of
$SU(3)$ and corresponds to a winding number $n$ [4,5]:
\begin{equation}
2\pi n=-\frac{i}{240\pi^2}\int_{S^5}dt d^4x\epsilon^{ijklm}
tr\left[U^{-1}\partial_iUU^{-1}\partial_jUU^{-1}\partial_kU
U^{-1}\partial_lUU^{-1}\partial_mU\right]\ \ .
\label{5.19}
\end{equation}

We can explore the significance of the above results. The action functional $W[A]$ is related
to the the vacuum functional (that is, the vacuum to vacuum amplitude) in the following way:
\begin{equation}
e^{iW[A]}=Z[A]=\lim_{\stackrel{\displaystyle{t_1 \rightarrow\  -\infty}}{\displaystyle{t_2\rightarrow\ +\infty}}}
\frac{\langle\Phi_o|e^{-iH(t_2-t_1)}|\Phi_o\rangle}%
{\langle\Phi_o|e^{-iH_o(t_2-t_1)}|\Phi_o\rangle}\ \ .
\label{5.20}
\end{equation}
If $A_\mu (x)$ becomes a pure gauge $g^{-1}\partial_\mu g$ in the vacuum, then the vacuum state $|\Phi_o\rangle$
can be characterized by this gauge. Similarly, if we consider $W[A^U]$, then the corresponding vacuum state
will be characterized by the gauge $U^{-1}g^{-1}\partial_\mu (gU)$. The vacuum state in the two cases can 
be different yielding the result
\begin{equation}
e^{iW[A]}=e^{i\Gamma[A,U]}\ e^{iW[A^U]}\ ,
\label{5.21}
\end{equation}
obtained by us earlier (Eq.(\ref{2.34})). If they do not differ, we have the non-anomalous case. This
situation is reminiscent of a Type II superconductor.
When a magnetic field pierces through it, the vacuum (i.e., the
superconductor itself) is characterized by a topological quantum number that specifies the quantized
magnetic flux enclosed by it. When no magnetic field pierces through it, we simply have a BCS vacuum.
One further remark: In building models of the nucleon,
clearly non-topological models of the kind proposed by Friedberg and Lee [25] should have an ordinary vacuum ($n=0$),
whereas topological soliton models of the kind proposed by Witten [4] should have a topologically nontrivial
vacuum. Phenomenological evidence from high energy elastic $pp \mbox{ and } \bar{p}p$ scattering, indicating a
superconducting type condensed quark-antiquark ground state forming the outer cloud of the nucleon [3,11],
appears to support strongly the case for the topological soliton model of the nucleon.
\section{Concluding Remarks}

Motivated by evidence from both low energy and high energy studies that
the nucleon is a topological soliton [1-3], we set out to examine how the (topological)
anomolous chiral action can be derived from the path-integral formalism. By generalizing
the Fujikawa approach, we obtained first an infinitesimal variation of the fermion measure
and then integrated it to derive formally the full anomalous action. This procedure requires
extending the gauge field to a five-dimensional space (the fifth dimension being a parameter
space), such that when the parameter $t=0$, we have the original gauge field $A_{\mu}(x)$,
but when $t=1$, we get an invariant gauge field $A_{\mu}^{U}(x)$. The expression for this anomalous 
action shows explicitly how it is obtained by integrating the anomaly (Eq.(2.35) ).

The anomaly, i.e., the covariant divergence of the current is given by an expression
similar to Fujikawa's:
$ D_\mu({\cal A})j^\mu(x,t)={\displaystyle
-\frac{1}{2}tr[\gamma^5\sum_n\varphi_n(x,t)\varphi_{n}^{\dagger}(x,t)]}$,
except that the gauge field and the wavefunctions depend on the parameter $t$ as well. Assuming the anomaly
to be consistent, we have shown in section 3, following the observation of Dunne and Trugenberger [14], that
the anomalous action can be identified as the Chern-Simons term. The question then becomes how to regularize
the theory, so that ${\displaystyle tr[\gamma^5\sum_n\varphi_n(x,t)\varphi_{n}^{\dagger}(x,t)]}$
yields the consistent anomaly.
To this end, we introduced Schwinger's proper-time representation for the Green's function and the 
gauge invariant point-splitting technique.
 We find that there are two ways of regularizing the theory -- both allowed by the WZ consistency
condition and the point-splitting technique. 
One way, indeed, leads to the consistent anomaly and therefore to a geometric theory with
Chern-Simons term. The other way leads to an anomaly free theory.

An alternative to the point-splitting method to derive the anomaly is the Pauli-Villars regularization
method. In the Abelian case, both methods yield the same result [20,21]. In the non-Abelian case, the
Pauli-Villars method has been mainly used to derive the anomaly [23,24]. Our work shows how the point-splitting
method can also be applied in the non-Abelian case. We note that even though the Fujikawa approach needs to
be generalized to connect with the Chern-Simons action, the Fujikawa regularization method (with appropriate
regulators) has been found to be equivalent to the Pauli-Villars regularization method for the axial anomaly
in four dimensions [26,27]. The same non-minimal terms appear in both methods, which can then be removed
by introducing additional counter terms in the action [26-28].

The usual discussion of the connection between the chiral anomaly and the $WZ$ consistency condition in
the path-integral framework is quite complicated, because: (i) the determinant of the Dirac operator
$\not \!\!{\cal D}$ can be specified in various ways, (ii) different specifications lead to different forms
for the Ward-Takahashi identities, (iii) different definitions of current operators are involved [29]. Our
strategy has been to require the chiral anomaly to be consistent from the beginning and see how the full 
anomalous action can be derived from the variation of the fermion measure.

We note that a low-energy effective action has been obtained by Balog [30] by integrating the non-Abelian
chiral anomalies. His results correspond to choosing $U(x,t)$ in Eq. (\ref{2.35}) as $exp\left[-t\Phi(x)\right]$ where
$\Phi(x)$ is the matrix-valued Goldstone field ($\Phi(x)=-2iT^a\phi^a(x)/f_{\pi},\ \phi^a(x)$'s are the pion
fields). Balog obtains not only the
expected geometrical part (an integral over the Bardeen anomaly in his case), but also other terms -- the 
``non-minimal" anomaly terms. Our calculations in sections 4 and 5 have shown that the expression for the anomaly
(Eq. (\ref{2.31})) needs to be regularized carefully, so that its evaluation yields the consistent anomaly,
and as such only
the term dictated by differential geometry. The non-minimal anomaly terms obtained by Balog come
from the regularization scheme of Andrianov and Bonora [31], which he used. 
As noted by Ebert and Reinhardt [32], the non-minimal anomaly terms can arise from the modulus
of the fermion determinant (more generally, from the gauge invariant part of the effective action, which
is $W[A^U]$ in our case). 
The non-minimal terms can be removed by adding
counter terms in the Lagrangian (as mentioned above), so that for a vector-axial-vector theory 
one is left only with the Bardeen anomaly.

In conclusion, our investigation establishes a bridge between the differential geometric
approach and the Fujikawa approach of obtaining the anomaly from the infinitesimal variation of the
fermion measure. This entails extending the gauge field to a five dimensional space
via a parameter dependent gauge transformation and requiring that the anomaly is consistent. The anomalous
action obtained in this way turns out to be the Chern-Simons term in five dimensions. We use the gauge-invariant
point-splitting technique and the WZ consistency condition to regularize the theory. We find that the
regularization can lead to an anomalous theory as well as to a non-anomalous theory.
Further investigation shows that  the nature of the vacuum determines whether the theory is anomalous or
non-anomalous.
\begin{center}
ACKNOWLEDGMENT
\end{center}

The authors wish to thank Gerald Dunne for his interest and comments. This work has been supported
in part by the U.S. Department of Energy under Grant DE-FG02-92ER40716.00.
\newpage
\appendix
\section{}
The WZ consistency condition plays an important role in our
investigation, because we require the anomaly $a({\cal A})$
to be consistent. Following the original work of Wess and Zumino
[33], the consistency condition is generally discussed in terms of
the generators of gauge transformations, which are functional
differential operators obeying the gauge algebra [13]. Here, we
discuss the condition in a way that makes its physical meaning 
transparent.

Under an infinitesimal gauge transformation
\begin{eqnarray}
A_\mu(x)\rightarrow A^{\theta}_\mu(x)&=&
e^{-\theta}A_\mu(x)e^\theta +e^{-\theta}\partial_\mu e^{\theta}
\nonumber \\
&=& A_\mu +D_\mu\theta\ \ . \label{A.1}
\end{eqnarray}
Correspondingly,
\begin{eqnarray}
W[A^\theta]-W[A] &=&\int\! d^4 x\left(D_\mu\theta\right)^a
\frac{\delta W}{\delta A^a_\mu} \nonumber \\
&=& -2i\int d^4 x\  tr\left[\theta a(A)\right]\ \ ,
\label{A.2}
\end{eqnarray}
where $a(A)=D_\mu j^\mu$, $j^{\mu,a}=i\frac{\delta W}%
{\delta A^a_\mu}$, $j^\mu=-iT^aj^{\mu,a}$, $\theta=-iT^a
\theta^a$. If we now consider the initial gauge configuration
to be $A^{\theta_1}_\mu$, then Eq. (\ref{A.2}) yields for an
infinitesimal gauge transformation $\theta_2$
\begin{equation}
W[A^{\theta_1\theta_2}]-W[A^{\theta_1}]=
-2i\int d^4 x\  tr\left[\theta_2 a(A^{\theta_1})\right]
\label{A.3}
\end{equation}
On the other hand, if the initial gauge field configuration
is $A_\mu(x)$,
\begin{equation}
W[A^{\theta_2}]-W[A]=
-2i\int d^4 x\  tr\left[\theta_2 a(A)\right]
\label{A.4}
\end{equation}
Subtracting (\ref{A.4}) from (\ref{A.3}), we find
\begin{equation}
W[A^{\theta_1\theta_2}]-W[A]-(W[A^{\theta_1}]-W[A])-(W[A^{\theta_2}]-W[A])
=-2i\int d^4 x\  tr\left[\theta_2\delta_{\theta_1}a(A)\right] \label{A.5}
\end{equation}
where $\delta_{\theta_1}a(A)=a(A^{\theta_1})-a(A)$. Now $A^{\theta_1\theta_2}=A^{\theta^\prime}$, where 
$e^{-\theta^\prime}=e^{-\theta_2}e^{-\theta_1}$, since two successive gauge transformations are equivalent
to one, and we have $\theta^{\prime}=\theta_1+\theta_2+\frac{1}{2}\left[\theta_1,\theta_2\right]+ ...$
From (\ref{A.5}), we therefore obtain
\begin{equation}
\int d^4 x\  tr\left[\theta^\prime a(A)\right]-\int d^4 x\  tr\left[\theta_1 a(A)\right]-\int d^4 x\  tr\left[\theta_2
a(A)\right]=\int d^4 x\  tr\left[\theta_2\delta_{\theta_1}a(A)\right].
\label{A.6}
\end{equation}
Let us next interchange $\theta_1$ and $\theta_2$. Then $A^{\theta_2\theta_1}=A^{\theta^{\prime\prime}}$, where
$e^{-\theta^{\prime\prime}}=e^{-\theta_1}e^{-\theta_2}$, and we get
\begin{equation}
\int d^4 x\  tr\left[\theta^{\prime\prime} a(A)\right]-\int d^4 x\  
tr\left[\theta_1 a(A)\right]-\int d^4 x\  tr\left[\theta_2
a(A)\right]=\int d^4 x\  tr\left[\theta_1\delta_{\theta_2}a(A)\right].
\label{A.7}
\end{equation}
Subtracting (\ref{A.7}) from (\ref{A.6}), we find
\begin{equation}
\int d^4 x\  tr\left[\ [\theta_1,\theta_2]a(A)\right]=\int\! d^4 x tr\left[\theta_2\delta_{\theta_1}a(A)
-\theta_1\delta_{\theta_2}a(A)\right],
\label{A.8}
\end{equation}
which is the $WZ$ consistency condition in the integrated form [6].

From (\ref{A.8}), it is easy to see that the covariant anomaly
$\tilde{a}(A)=\frac{1}{64\pi^2}\epsilon^{\lambda\mu\rho\sigma}G_{\lambda\mu}(A)G_{\rho\sigma}(A)$
does not satisfy the consistency condition. Since $G_{\lambda\mu}(A^{\theta})=e^{-\theta}G_{\lambda\mu}(A)e^{\theta}$, 
$\delta_{\theta}\tilde{a}(A)=\left[\tilde{a}(A),\theta\right]$ and the RHS of (\ref{A.8}) for the covariant anomaly
is
\begin{equation}
\int d^4 x\  tr\left[\theta_2\left[\tilde{a}(A),\theta_1\right]
-\theta_1\left[\tilde{a}(A),\theta_2\right]\right]=2\int d^4 x\  tr\left[\ [\theta_1,\theta_2]\tilde{a}(A)\right],
\label{A.9}
\end{equation}
which is twice the corresponding LHS.

We now explicitly show that the consistent anomaly 
\[
\frac{1}{48\pi^2}\epsilon^{\lambda\mu\rho\sigma}\partial_\lambda\left[{\cal A}_\mu
\partial_\rho{\cal A}_\sigma + \frac{1}{2} {\cal A}_\mu{\cal A}_\rho{\cal A}_\sigma \right]
\]
satisfies the consistency equation (\ref{A.8}). For this purpose, we introduce the form notation: 
\begin{eqnarray}
a(A)&=& \frac{1}{48\pi^2}d\left[AdA+\frac{1}{2}A^3\right] \nonumber \\
&=& \frac{1}{96\pi^2}d\left[ AdA+ dAA+A^3\right], \label{A.10}
\end{eqnarray}
and for an infinitesimal gauge transformation
\begin{eqnarray}
\delta_{\theta}A&=& A^\theta-A \nonumber \\
&=& d\theta+[A,\theta].
\label{A.11}
\end{eqnarray}
Eq. (\ref{A.10}) yields
\begin{equation}
\delta_\theta a(A)=d\left\{[K(A),\theta]-\frac{1}{96\pi^2}Ad\theta A\right\},
\label{A.12}
\end{equation}
where
\begin{equation}
K(A)=\frac{1}{96\pi^2}\left( AdA+ dAA+A^3\right).
\label{A.13}
\end{equation}
Therefore,
\begin{eqnarray}
\int tr\ \theta_2\delta_{\theta_1} a(A)&=& \int tr\ \theta_2d[K,\theta_1]
 -\frac{1}{96\pi^2}\int tr\ \theta_2d(Ad\theta_1 A) \nonumber \\
&=& \int tr\ d\theta_2[\theta_1,K ]+\frac{1}{96\pi^2}\int tr\ d\theta_2 Ad\theta_1 A\ .
\label{A.14}
\end{eqnarray}
Interchanging $\theta_1$ and $\theta_2$, 
\begin{eqnarray}
\int tr\ \theta_1\delta_{\theta_2} a(A)&=&\int tr\ d\theta_1[\theta_2,K]+\frac{1}{96\pi^2}\int tr\ d\theta_1
Ad\theta_2 A \nonumber \\
&=&\int tr\ d\theta_1[\theta_2,K]+\frac{1}{96\pi^2}\int tr\ d\theta_2 Ad\theta_1 A \ .
\label{A.15}
\end{eqnarray}
Subtracting (\ref{A.15}) from (\ref{A.14}), we find
\begin{eqnarray}
\int tr\left\{\theta_2\delta_{\theta_1} a(A)-\theta_1\delta_{\theta_2} a(A)\right\}&=&
\int tr[d\theta_2,\theta_1]K+\int tr [\theta_2,d\theta_1]K \nonumber \\
&=& \int tr\ d[\theta_2,\theta_1]K \nonumber \\
&=&\int tr [\theta_1,\theta_2]dK \nonumber \\
&=& \int tr [\theta_1,\theta_2]a(A),
\label{A.16}
\end{eqnarray}
which is the consistency condition (\ref{A.8}).


\newpage
\begin{references} 
\bibitem{1} I. Zahed and G.E. Brown, Phys.Reports {\bf 142} (1986) 1.
\bibitem{2} R.K. Bhaduri, {\em Models of The Nucleon: From Quarks to Soliton} (Addison-Wesley Publishing Co.,
1988).
\bibitem{3} M.M. Islam, in {\em Proceedings of the Workshop on Quantum Infrared Physics}, edited by
H.M. Fried and B. Muller (World Scientific, 1995),p.401.
\bibitem{4} E. Witten, Nucl. Phys. {\bf B 223} (1983) 422, 433.
\bibitem{5} P. Goddard and P. Mansfield, Rep. Prog. Phys. {\bf 49} (1986) 725.
\bibitem{6} B. Zumino, Y.-S. Wu, and A. Zee, Nucl. Phys {\bf B 239} (1984) 477.
\bibitem{7} B. Zumino, in {\em Current Algebra and Anomalies}, ed. by S.B. Treiman et al. (World Scientific,
1985).
\bibitem{8} J.L. Petersen, Acta Physica Polonica, {\bf B 16} (1985) 271.
\bibitem{9} K. Fujikawa, Phys. Rev. {\bf D 21} (1980) 2848; {\bf 22} (1980) 1499 (E).
\bibitem{10} K. Fujikawa, Phys. Rev. {\bf D 29} (1984) 285.
\bibitem{11} M.M. Islam, Z. Phys. {\bf C 53} (1992) 253.
\bibitem{12} W.A. Bardeen and B. Zumino, Nucl. Phys. {\bf B 244} (1984) 421.
\bibitem{13} R.D. Ball, Phys.Reports {\bf 182} (1989) 1.
\bibitem{14} G.V. Dunne and C.A. Trugenberger, Ann. Phys. (N.Y.) {\bf 204} (1990) 281.
\bibitem{15} J. Schwinger, Phys. Rev. {\bf 82} (1951) 664.
\bibitem{16} This, of course, means that in a free theory the anomaly must vanish. Similar subtraction
has been done by Banerjee et al. in establishing the relation between the consistent current and the covariant
current: H. Banerjee, R. Banerjee, and P. Mitra, Z. Phys. {\bf C 32} (1986) 445.
\bibitem{17} Formally, both $G_R (x,y)$ and $G_L (x,y)$ are equal to $\displaystyle{\sum_n\frac{\varphi_n
(x)\varphi^{\dag}_n(y)}{\lambda_n^2+m^2}}$.
\bibitem{18} This result can be easily seen by noticing that the infinite sum on the RHS of (\ref{4.11})
leads to the formal expression:
\[
G_R(x,y)= \int\frac{d^4k}{(2\pi)^4}\ e^{ik\cdot(x-y)}\frac{1}{m^2-k^2+{\cal D}^2+2ik\cdot{\cal D}
+\frac{1}{2}\gamma^\rho\gamma^\sigma G_{\rho\sigma}(x)}.
\]
\bibitem{19} R. Jackiw and K.A. Johnson, Phys. Rev. {\bf 182}
(1969) 1459.
\bibitem{20} R. Jackiw, in {\em Current Algebra and Anomalies}, ed. by S.B. Treiman et al. (World Scientific,
1985).
\bibitem{21} P.D.B. Collins, A.D. Martin, and E.J. Squires, {\em Particle Physics and Cosmology} (John Wiley  and
Sons, 1989), Ch. 5.
\bibitem{22} If a wavefunction $\varphi(x)$ transforms under a gauge transformation to
$\varphi^{\prime}(x)=U^{-1}(x)\varphi(x)$, then the combination $\varphi^{\dag}(y)\Omega (y,x)\varphi(x)$,
just like $\varphi^{\dag}(x)\varphi(x)$, is gauge invariant, because under the gauge transformation
\[
\Omega(y,x)\rightarrow \Omega^{\prime}(y,x)=U^{-1}(y)\Omega(y,x)U(x).
\]
\bibitem{23} W.A. Bardeen, Phys. Rev. {\bf 184} (1969) 1848.
\bibitem{24} D. Gross and R. Jackiw, Phys.Rev. {\bf D 6} (1972) 477.
\bibitem{25} T.D. Lee, {\em Particle Physics and Introduction to Field Theory} (Harwood Academic Publishers,
1981), Ch. 20.
\bibitem{26} M.B. Einhorn and D.R.T. Jones, Phys. Rev. {\bf D 29} (1984) 331.
\bibitem{27} S.-K. Hu, B.-L. Young, and D.W. Mckay, Phys. Rev. {\bf D 30} (1984) 836.
\bibitem{28} A.P. Balachandran, G. Marmo, V. P. Nair, and C.G. Trahern, Phys. Rev. {\bf D 25} (1982) 2713.
\bibitem{29} K. Fujikawa, Phys. Rev. {\bf D 31} (1985) 341.
\bibitem{30} J. Balog, Phys. Lett. {\bf 149 B} (1984) 197.
\bibitem{31} A. Andrianov and L. Bonora, Nucl. Phys. {\bf B 233} (1984) 232,247.
\bibitem{32} D. Ebert and H. Reinhardt, Nucl. Phys. {\bf B 271} (1986) 188.
\bibitem{33} J. Wess and B. Zumino, Phys. Lett. {\bf 37B }(1971) 95.
\end{references}
\end{document}